\documentclass{aa}  
\usepackage{graphicx}
\usepackage{color}
\usepackage{url}
\usepackage{hyperref}
\usepackage[normalem]{ulem}
\usepackage{txfonts}
\usepackage{natbib}
\newcommand{\psrtar}{Swift\,J1756.9$-$2508}

\newcommand{\gtap}{\mathrel{\hbox{\rlap{\lower.55ex \hbox {$\sim$}}
                   \kern-.3em \raise.4ex \hbox{$>$}}}}
\newcommand{\ltap}{\mathrel{\hbox{\rlap{\lower.55ex \hbox {$\sim$}}
                   \kern-.3em \raise.4ex \hbox{$<$}}}}

\newcommand{\nustar}{{\it NuSTAR\/}}
\newcommand{\nicer}{{NICER\/}} %not a satellite, not in italic   

\newcommand{\xmm}{{\it XMM-Newton\/}}
\newcommand{\swift}{{\it Swift\/}}
\newcommand{\Integ}{{INTEGRAL\/}} %a satellite, but not named, not in italic according to A&A rules
\newcommand{\rxte}{{\it RXTE\/}}
\newcommand{\hxmt}{{{\it Insight}-HXMT\/}}

\def\compps{{\sc compps}}

\def\be{\begin{equation}} 
\def\ee{\end{equation}}

\begin{document}

% suggestion for slightly shorter title - dkg
   \title{% X-ray and soft \gr\ 
   Broadband X-ray spectra and timing of the accreting millisecond pulsar \psrtar\ during its 2018 and 2019 outbursts}
   \titlerunning{High-energy characteristics of \psrtar}
   \authorrunning{Z. Li et al.}

  % \subtitle{An XMM-Newton, NICER, NuSTAR and INTEGRAL view\\ of the 0.3--300 keV X-ray band}

   \author{Z. S. Li\inst{1}
           \and
              L. Kuiper\inst{2}
            \and
      M. Falanga\inst{3}
             \and
    J. Poutanen\inst{4,5,6}
    \and
    S. S. Tsygankov\inst{4,5}
    \and
    D. K. Galloway\inst{7,8}
    \and
        E. Bozzo\inst{9}
    \and
     Y. Y. Pan\inst{1}
     \and
    Y. Huang\inst{10}
    \and
    S. N. Zhang\inst{10}
    \and
    S. Zhang\inst{10}
          }
   \offprints{Z. Li}

   \institute{Key Laboratory of Stars and Interstellar Medium, Xiangtan University, Xiangtan 411105, Hunan, P.R. China\\
              \email{lizhaosheng@xtu.edu.cn}
               \and
            SRON-Netherlands Institute for Space Research, Sorbonnelaan 2,  3584 CA, Utrecht, The Netherlands    
              \and
              International Space Science Institute (ISSI), Hallerstrasse 6, 3012 Bern, Switzerland
              \and
              Department of Physics and Astronomy, University of Turku, FI-20014, Finland
              \and
              Space Research Institute of the Russian Academy of Sciences, Profsoyuznaya str. 84/32, 117997 Moscow, Russia
              \and
              Nordita, KTH Royal Institute of Technology and Stockholm University, Roslagstullsbacken 23, SE-10691 Stockholm, Sweden
              \and
              School of Physics and Astronomy, Monash University, Australia, VIC 3800, Australia
              \and
              OzGRav-Monash, School of Physics and Astronomy, Monash University, Victoria 3800, Australia
              \and
             University of Geneva, Department of Astronomy, Chemin d'Ecogia 16, 1290, Versoix, Switzerland
              \and
              Key Laboratory of Particle Astrophysics, Institute of High Energy Physics, Chinese Academy of Sciences, 19B Yuquan Road, Beijing 100049, China
              }

   \date{Received xx  / Accepted xx}

% \abstract{}{}{}{}{} 
% 5 {} token are mandatory
  \abstract{
The accreting millisecond X-ray pulsar \psrtar\ launched into an outburst in April 2018 and June 2019 -- 8.7~years after the previous period of activity. We investigated the temporal, timing, and spectral properties of these two outbursts 
% across 1--250 keV band by 
using data from \nicer, \xmm, \nustar, \Integ, \swift,\ and \hxmt. The two outbursts exhibited similar 
% phenomena, i.e., the 
broadband spectra and X-ray pulse profiles.
% and the outburst profiles. % arguably not, as discussed & shown in Fig 3 --- dkg
For the first time, we report the detection of the pulsed emission up to $\sim100$ keV that was observed by \hxmt\ during the 2018 outburst. We also found the pulsation up to $\sim60$ keV that was observed by \nicer\ and \nustar\ during the 2019 outburst.  We performed a coherent timing analysis combining the data from the two outbursts. The binary system is well described by a constant orbital period over a time span of $\sim12$ years. 
%
% I would suggest omitting this sentence if required for space --- dkg
The time-averaged  broadband spectra are well fitted by the absorbed thermal Comptonization model \compps\  in a slab geometry with an electron temperature, $kT_{\rm e}=40$--50 keV, Thomson optical depth $\tau\sim 1.3$, blackbody seed photon temperature $kT_{\rm bb,seed}\sim $0.7--0.8~keV, and hydrogen column density of $N_{\rm H}\sim 4.2\times10^{22}$~cm$^{-2}$. 
% added by dkg
We searched the available data for type-I (thermonuclear) X-ray bursts, but found none, which is unsurprising given the estimated low peak accretion rate ($\approx0.05$ of the Eddington rate) and generally low expected burst rates for hydrogen-poor fuel.
Based on the history of four outbursts to date, we estimate the long-term average accretion rate at roughly $5\times10^{-12}\ M_\odot\,{\rm yr}^{-1}$ for an assumed distance of 8~kpc. The expected mass transfer rate driven by gravitational radiation in the binary implies the source may be no closer than 4~kpc.
% The reason of non-detection of type-I X-ray bursts from \psrtar\ has been discussed. 
%
\psrtar\ is the third low mass X-ray binary exhibiting ``double'' outbursts, which are separated by much shorter intervals than what we typically see and are likely to result from interruption of the accretion flow from the disk onto the neutron star. Such behavior may have important implications for the disk instability model.
}
  % context heading (optional), leave it empty if necessary  
%   {\psrtar\ }
%  % results heading (mandatory)
   
   \maketitle
%
%________________________________________________________________

\section{Introduction}
\label{sec:intro}

\psrtar\ was discovered by \swift-BAT during its 2007 outburst \citep{Krimm07a, Linares08}.  Coherent X-ray pulsations at a frequency of $\sim$182 Hz confirmed the compact object to be an accreting millisecond X-ray pulsar (AMXP) \citep{Markwardt07}. In the following observations, carried out by \swift\ and \rxte, an orbital period of 54.7 min was measured and, based on its mass function, the mass of the companion star -- a highly evolved white dwarf -- was determined to be in the range of $0.0067-0.03 M_\sun$ \citep{krimm2007}. 

The AMXPs are rapidly spinning, old, recycled neutron stars (NSs) hosted in low-mass X-ray binaries (LMXB). As a binary system evolves through phases of accretion onto the NS, it gains angular momentum from the accreted material, which is sufficient to spin-up the NS to a rotation period equilibrium in the millisecond range, that is, the recycling of old radio-pulsars to millisecond periods \citep{Alpar82,Wijnands98,falanga05,papitto13c}. On the other hand, between the outbursts, long-term monitoring shows some AMXPs to exhibit spin-down in quiescence \citep[see, e.g.,][]{Hartman2009,Patruno10,Papitto10}. For reviews of the properties of these objects, we refer to \citet{Wijnands2006}, \citet{Poutanen06}, \citet{Patruno2012}, \citet{Campana18}, \citet{papitto20}, and \citet{Salvo20}. 

The source, \psrtar, was shown to have undergone another three outbursts: in 2009 \citep{Patruno09}, 2018 \citep{Sanna18,Bult18,Rai19,Atel11566}, and 2019 \citep{ATel12882}.  For the first three recurrent outbursts between 2007 and 2018, there was no orbital period variation detected \citep{krimm2007,Patruno10,Bult18,Sanna18}. Moreover,  the upper limit on the orbital period derivative  is $|\dot{P}_{\rm orb}|<7.4\times10^{-13}~{\rm s~s^{-1}}$ from  the total 11-year span of the observations, which is consistent with the prediction of a conservative mass transfer in the binary system \citep{Bult18}. \psrtar\ exhibited long-term spin-down behavior with a spin frequency derivative of $|\dot{\nu}|\lesssim 3\times 10^{-13}~{\rm Hz~s^{-1}}$, measured by \citet{Patruno10},   $|\dot{\nu}|\sim7.3\times10^{-16}~{\rm Hz~s^{-1}}$  by \citet{Bult18} and a  smaller value $|\dot{\nu}|\sim4.8\times10^{-16}~{\rm Hz~s^{-1}}$ by \citet{Sanna18}. Assuming that the rotational energy loss due  to the  magnetic dipole emission dominated the spin evolution of \psrtar, the magnetic field strength at the stellar magnetic poles was constrained to the range of $(3-6)\times10^8~\rm{G}$ \citep[see, e.g.,][and references therein]{Patruno10,Bult18,Sanna18}.

Most NSs hosted in LMXB systems exhibit type I X-ray bursts 
% \citep{liu07}. 
% a more up-to-date list is here --- dkg
\citep{Galloway20}. 
Type I bursts are thermonuclear explosions on the surface of the accreting NS, triggered by unstable hydrogen or helium burning. They are typically characterized by a fast rise time of $\sim 1-2$ s, followed by exponential-like decays, as well a  gradual softening due to the cooling of the NS photosphere, with durations ranging from seconds to minutes and recurrence times from a few hours to days \citep[see, e.g.,][for reviews]{Lewin93,SB06,Galloway21}. The burst spectra are described by a blackbody with peak temperatures reaching $kT_{\rm bb}\approx 3$ keV, where the luminosity can reach the Eddington limit, $L_{\rm Edd}\approx 2\times10^{38}$ erg s$^{-1}$, and the total burst energy release is on the order of $\sim 10^{39}$ erg \citep[see][for a review]{Lewin93,SB06,Galloway08b}. Several thousand bursts have been observed to date \citep[see, e.g.,][]{Galloway08b,Galloway20}. However, 7 sources (including \psrtar) out of 21 known AMXPs, including 20 AMXPs mentioned in Table 1 from \citet{Salvo20} as well as the newly confirmed AMXP IGR J17494$-$3030 \citep{Ng21}, have not shown type-I X-ray bursts during their outbursts. Hence, the distance to  \psrtar\ is still unknown. 
Below, for a number of estimates, we assume a distance of 8 kpc (since the source lies close to the direction to the Galactic center). 

In this paper, we report on the broadband (1--300 keV) spectral and timing results using the available  high-energy data, including \Integ\ ($20-300$ keV), two \nustar\ ($3-79$ keV), one \xmm\ ($1-10$ keV), eight \nicer\ ($0.2-12$ keV), nine \swift,\  and five \hxmt\ ($5-250$ keV) observations during the 2018 outburst from \psrtar, as well as one \nustar, one \swift,\  and seven \nicer\ observations during its 2019 outburst. All the monitoring observations  have been applied to shed light on the physical processes acting upon \psrtar.

\section{Observations and data reduction}
\label{sec:observation}

We reduce the data from \psrtar\ collected during its 2018 and 2019 outbursts.
For the 2018 outburst, \Integ\ was the first to detect  the enhanced X-ray emission from \psrtar\ \citep{ATel11497}. Several high-energy facilities carried out the follow-up observations covering the whole outburst \citep{ATel11502, ATel11505, ATel11523,ATel11581,ATel11603}.
%, see Fig.~\ref{fig:outburst_profile}. \red{JP: reference  to Fig.2 comes before that to Fig.1. I suggest to refer to Fig. 2 later when the details of the observations are given. Alternatively, Fig.2 should be Fig.1 and in the caption one can refer to spectral modeling section where details of the fluxes determination are given } 
During the 2019 outburst, \nicer, \nustar, and \swift\ observed \psrtar\ \citep{ATel12882}. Below, we introduce the data reduction part for all instruments. In addition, we provide in Table~\ref{table:observations} a log of all observations during the 2018 and 2019 outbursts. We adopted Solar System ephemeris DE405 and the X-ray position of \psrtar\ \citep{krimm2007} to apply the barycentering corrections for the timing analysis using \Integ\ (Sect.~\ref{sec:integral}), \nicer\ (Sect.~\ref{sec:nicer}),  \nustar\ (Sect.~\ref{sec:nustar}), \xmm\ (Sect.~\ref{sec:xmm}), and \hxmt\ (Sect.~\ref{sec:hxmt}) observations.

%=======================
\begin{table}%[h] 
{\small
\caption{Observations of \psrtar\ for the 2018 and 2019 outbursts, respectively. 
%\red{*It makes sense to add dates of the observations*}
}
\centering
\begin{tabular}{lcccc} 
\hline 
Mission & Obs. ID & Instrument & Date\tablefootmark{a} & Exposure \\
 &  & & (d)  &(ks) \\
\hline 
\noalign{\smallskip}  
\multicolumn{5}{c}{2018 outburst} \\
\nustar       &  90402313002    & FPMA/FPMB &  8.36   &39.5\\ 
              &  90402313004    & FPMA/FPMB &  14.13   & 61.0\\ 
\nicer        &  1050230101     & XTI &  3.64    &7.5  \\ 
              &  1050230102     & XTI &  4.61   & 7.1 \\ 
              &  1050230103     & XTI &  7.89   & 2.4 \\ 
              &  1050230104     & XTI &  8.02    & 10.7 \\ 
              &  1050230105     & XTI &  9.33    & 4.1 \\ 
              &  1050230106     & XTI &  10.02    & 6.2 \\ 
              &  1050230107     & XTI &  11.06    & 5.0 \\ 
              &  1050230108     & XTI &  25.85     & 1.5 \\ 
\xmm          &  0830190401     & RGS/MOS/PN  & 8.09  &       66   \\
\Integ        &   1936, 1937       & ISGRI  & -1.23, 1.42 & 129 \\ 
              &   1939 (ToO)         & ISGRI &   7.68      & 85 \\ 
              &   1940   & ISGRI  &  10.01  &   \\ 
              &   1942   & ISGRI  &   14.91 & 190\tablefootmark{b} \\ 
              &   1944   & ISGRI  &   20.20 &  \\ 
\hxmt         &   P011469200101   &  ME \& HE &   5.13     & 4.3  \\
              &   P011469200102   &  ME \& HE  &  5.33     & 1.1  \\
              &   P011469200103   &  ME \& HE  &  5.47     & 0.3 \\
              &   P011469200104   &  ME \& HE &   5.60     & 1.2 \\
              &   P011469200105   &  ME \& HE  &  5.71     & 5.9 \\
\swift        &   00030952018     &   XRT     &  2.67      & 1.0 \\ 
              &   00088662001     &   XRT     &  8.77      & 2.2 \\ 
              &   00030952019     &   XRT     &  9.78      & 1.0 \\ 
              &   00030952020     &   XRT     &  10.38      & 0.4 \\ 
              &   00030952021     &   XRT     &  12.59      & 1.3 \\ 
              &   00088662002     &   XRT     &  14.82      & 2.1 \\ 
              &   00030952022     &   XRT     &  18.75      & 1.0 \\ 
              &   00030952023     &   XRT     &  24.59      & 1.2 \\ 
              &   00030952024     &   XRT     &  27.44      & 1.4 \\ 

\noalign{\smallskip}  
\hline  
\multicolumn{5}{c}{2019 outburst} \\
%\hline 
\noalign{\smallskip}  
\nustar               & 90501329001   & FPMA/FPMB &   11.85    & 37.7\\ 
\nicer                      &  2050230101     & XTI  & 9.10    & 1.5  \\ 
                            &  2050230102     & XTI &  9.56    & 3.6 \\ 
                            &  2050230103     & XTI &  11.10    & 5.4 \\ 
                            &  2050230104     & XTI &  11.50     & 11.7 \\ 
                            &  2050230105     & XTI &  16.51    & 1.1 \\ 
                            &  2050230106     & XTI &  19.02    & 1.7 \\ 
                            &  2050230107     & XTI &  21.16    & 1.1 \\                             

\swift   &   00030952025     &   XRT     &   10.53      & 0.9  \\ 
\noalign{\smallskip}  
\hline  
\end{tabular}  
\tablefoot{ \tablefoottext{a}{The start time of the observations since MJD 58208 and 58644.5 for the 2018 and 2019 outbursts, respectively.}
\tablefoottext{b}{The total exposure time of Rev. 1940, 1942, and 1944.}
}
%            \tablefoottext{b}{Unabsorbed flux in the 0.3 -- 150 keV energy range.}
%          }
\label{table:observations} 
}
\end{table} 

%%%%%%%%%%%%%%%%%%%%%%%% Fig-1 %%%%%%%%%%%%%%%%%%%%%%%%
\begin{figure}[h]
\centering
%\vspace{-3cm}
%\hspace{-1.65cm}
  \includegraphics[width=9cm,bb=65 160 575 595,clip= ]{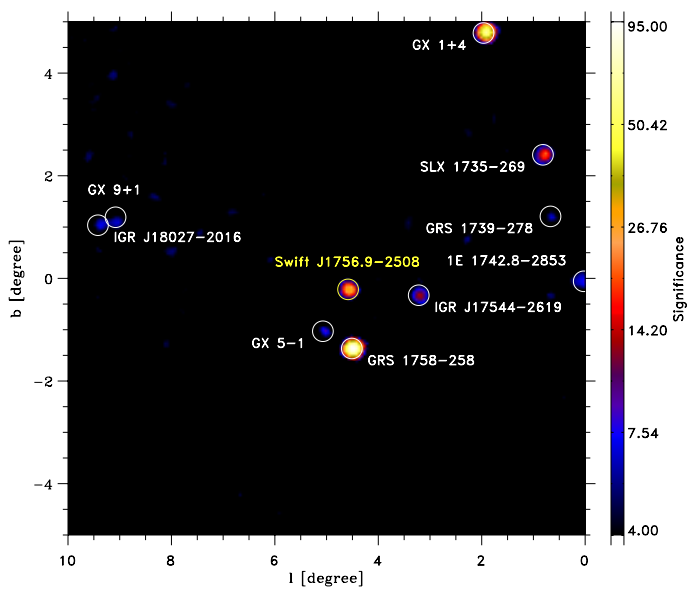}
% \vspace{-3.0cm}
  \caption{INTEGRAL-ISGRI sky image in the 20--50 keV band (significance map, in Galactic coordinates ($l,b$)), of the field of view around the AMXP \psrtar\ taken during time interval MJD 58209.35--58219.50  with an effective exposure time of 248 ks (rev. 1937, 1939, and 1940 combined). The source was detected with a significance of $\sim30\sigma$.
 }
  \label{fig:isgri_map}
\end{figure}
%%%%%%%%%%%%%%%%%%%%%%%% Fig-1 %%%%%%%%%%%%%%%%%%%%%%%%

\subsection{INTEGRAL}
\label{instr_integral}
\label{sec:integral}

Our \Integ\ \citep{w03} data set comprises all the observations covering the 2018 outburst. It consists of 160 stable pointings (science windows, ScWs) with a source position offset from the center of the field of view $\lesssim 12\fdg0$. The different satellite pointings in the direction of \psrtar, each lasting  $\sim 2-3$\,ks, were performed between March 29 and April 20, 2018. Specifically, the satellite revolutions included in the analysis were: 1936--1937, 1939--1940, 1942, and 1944. This includes also a dedicated Target of Opportunity (ToO) observation during revolution 1939 beginning on April 7  (MJD 58215.81170).
%for a total effective exposure time of 85~ks on the source. 
The data reduction was performed using the standard Offline Science Analysis (OSA\footnote{http://www.isdc.unige.ch/integral/analysis}) version 10.2, distributed by the Integral Science Data Center \citep{c03}. The algorithms used for the spatial and spectral analysis are described in \citet{gold03}. We analyzed data from the IBIS-ISGRI coded mask telescope \citep{u03,lebrun03}, at energies between 22 and 300 keV, and from the JEM-X monitor, module 1 and 2 \citep{lund03} between 5 and 20 keV.  Because most of the pointings were not aimed at \psrtar\ as the primary target, the source was rarely within the JEM-X field of view of $<3\fdg5$. Therefore, we did not use the JEM-X data for the spectral analysis.

We show in Fig. \ref{fig:isgri_map} part of the ISGRI field of view for 20--50 keV energy range (significance map) centered on the position of \psrtar. The source was clearly detected in the mosaic at a significance of $\sim30\sigma$ in the 20--50 keV energy range and still significant with $\sim5.8\sigma$ in the higher 100--150 keV energy range. The best-determined position is at $\alpha_{\rm J2000} = 17^{\rm h}56^{\rm
m}57\fs35$ and $\delta_{\rm J2000} = -25{\degr}06\arcmin27\farcs8$, with an associated uncertainty
of $0\farcm5$ at the 90\% c.l. (20--50 keV; \citealt{gros03}).

We extracted the IBIS-ISGRI light curve of \psrtar\ at the resolution of one ScW for the entire observational period covered by \Integ\ (see Sect.~\ref{sec:outburst_profile}). However, the ISGRI spectra were extracted only using the dedicated continuous ToO observation in revolution 1939, as these occurred nearly simultaneously with the \nicer\, \nustar\, and \xmm\ observations (see Sect.~\ref{sec:nicer}, \ref{sec:nustar}, \ref{sec:xmm}) and permitted the most accurate description of the source averaged broadband high-energy emission. The outburst profile is described in Sect. \ref{sec:outburst_profile} and the averaged broadband spectrum of the source, as it was measured almost simultaneously by all these instruments, is described in Sect.~\ref{sec:spectrum}. We note that \Integ\ did not carry out observations of \psrtar\ during its 2019 outburst.

%We describe in Sect.~\ref{sec:timing} the results of the source pulse profile analysis in the hard X-rays as obtained from the ISGRI event files. This is compared to the pulse profiles obtained in the soft X-rays with \nicer\, \nustar\, \xmm\, and Insight-HXMT.  

%To search for X-ray bursts, the ISGRI lightcurves are calculated from events selected according
%to the detector illumination pattern for Swift J1756-. For ISGRI we used an
%illumination factor threshold of 0.6 for the energy range 18--40 keV;
%for JEM-X we used the event list of the whole detector in the 3--20
%keV energy band.

%Using all available ISGRI data and publically available Swift BAT daily averaged (15-50 keV) data I produced an %outburst profile for Swift J1756.9-2508 during its April-2018 outburst. The result (ISGRI-BAT combined) is %attached. Max. flux reached is about 60 mCrab in 20-50 keV band. 

%Emission is clearly detected well above (> 79 keV) the NuSTAR energy band:  100-150 keV 5.8 sigma,
%and its spectrum - outburst averaged - is very Crab like with fluxes of 20.97 +/- 0.71 mCrab,
%21.54 +/- 1.60 mCrab and 17.26 +/- 3.54 mCrab in the 20-50, 50-100 and 100-150 keV bands, respectively.
%The Crab obs. during Rev-1943 has been used to convert the count rates to fluxes.

%%%%%%%%%%%%%%%%%%%%%%%% Fig-2 %%%%%%%%%%%%%%%%%%%%%%%%
\begin{figure} %[t]
\centering
\includegraphics[width=8.5cm]{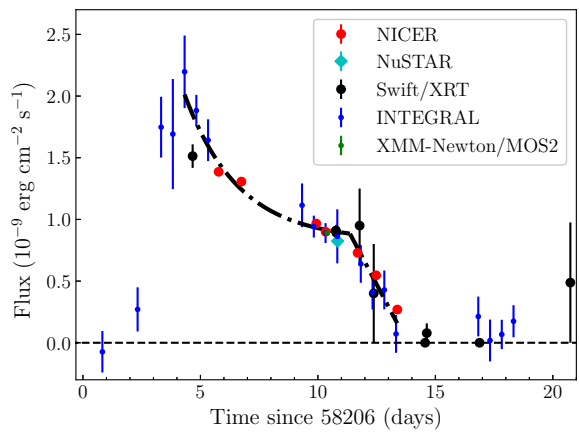}
\includegraphics[width=8.5cm]{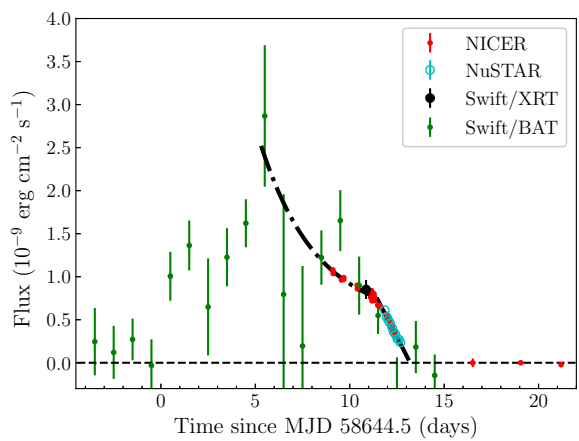}
\caption{Evolution of the bolometric fluxes (0.1--300 keV)  of \psrtar\ for the 2018 (top panel) and 2019 (bottom panel) outbursts; see Sect.~\ref{sec:outburst_profile} for more details.  The dot-dash  lines correspond to the best-fit exponential profile, $F \propto e^{-t/2.3~{\rm d}}$ and $F \propto e^{-t/3~{\rm d}}$, followed by a linear decay, respectively.
}
\label{fig:outburst_profile}
\end{figure}
%%%%%%%%%%%%%%%%%%%%%%%% Fig-2 %%%%%%%%%%%%%%%%%%%%%%%%

\subsection{NICER}
\label{sec:nicer}

The Neutron star Interior Composition ExploreR (\nicer), launched on June 3, 2017, is an International Space Station payload dedicated to (spectral) timing studies
in the 0.2--12 keV band at an unprecedented time resolution of $\sim 100$ ns  \citep{arzoumanian2014}.  During the 2018 outburst, \nicer\ started regular observations of \psrtar\ on April  3, 2018 (MJD 58211.6; obs. ID 1050230101), namely, a few days after the detection by \Integ,\, and it ended on 
April 25, 2018 (MJD 58233.8; obs. ID 1050230108) when the source reached undetectable levels. During the 2019 outburst, \nicer\ observed \psrtar\ seven times, including obs. ID 2050230101 through 2050230107.  

We carried out standard data processing using the NICER Data Analysis Software (NICERDAS). The default filtering criteria were applied to extract the cleaned event data. The spectra and light curves were obtained by {\tt xselect}.  The  spectrum from obs. ID 1050230104 was used together with those from the other instruments in the 2018 outburst. The background spectrum is adopted from obs. ID 1050230107, when \psrtar\ was in a quiescent state with negligible X-ray emission.
In the pulse-profile analysis of the 2018 \nicer\ data (see Sect. \ref{sec:pulse_time_lag}) we used obs. ID 1050230108 for the estimation of the background to obtain background corrected fractional amplitudes.

For the 2019 outburst, the \nicer\ observation with obs. ID 2050230104 overlapped with the (single) \nustar\ observation. In order to obtain a simultaneous broadband spectrum, the data between MJD 58656.3485 and 58656.9060 were analyzed; in this case, the background spectrum was from obs. ID 2050230107.  The redistribution matrix file and ancillary response file were taken from the official webpage.\footnote{\url{https://heasarc.gsfc.nasa.gov/docs/nicer/tools/nicer_bkg_est_tools.html}} For details on the timing analysis of \nicer\ data we refer to Sect. \ref{sec:timing}.
% \textbf{For the timing analysis, }  

%The last three observations in the 2019 outburst showed \psrtar\ back into quiescent state.
\subsection{NuSTAR}
\label{sec:nustar}

\nustar\ observed \psrtar\ on April 8, 2018  (obs. ID 90402313002; MJD $\sim 58216.4$), April 14, 2018  (obs. ID 90402313004; MJD $\sim58222.1$), and on June 22, 2019 (obs. ID 90501329001; MJD $\sim58656.3$). The second observation during the 2018 outburst was carried out when \psrtar\ returned into quiescent state, so it was not considered in this work. For the first 2018 observation, we cleaned the event file using the  \nustar\ pipeline tool {\tt nupipeline} for both FPMA and FPMB. The light curves and spectra were extracted from a circular region with a radius of  $60\arcsec$ centered on the location of \psrtar\ (source region) by using {\tt nuproducts}, and the response files were produced simultaneously. 
To extract the background spectra, we chose a source-free circular background region located on the same chip with a radial aperture of $60\arcsec$.

In the timing analysis, we barycentered event data from the source region using {\tt HEASOFT} multi-mission tool {\tt barycorr v2.1,} with \nustar\ fine clock-correction file \#110, yielding time tags accurate at 60--100~$\mu$s level \citep[][]{Bachetti2021}. To obtain background corrected timing characteristics, such as fractional amplitudes, we used the above mentioned background region. For the observation carried out during the 2019 outburst, similar procedures were applied as for the 2018 outburst, except that the spectra were cut in the same interval as the \nicer\  obs. ID 2050230104. For more details, see Sect.~\ref{sec:nicer}.

\subsection{XMM-Newton}
\label{sec:xmm}

On April 8, 2018, \xmm\ started  an observation of \psrtar\ (MJD 58216.07). 
The (imaging) EPIC-pn instrument \citep[][0.15--12 keV]{struder2001} on board \xmm\ was operated in timing mode (TM), 
allowing for timing studies at a $\sim 30 ~\mu$s time resolution. The other (imaging) EPIC instrument equipped with two cameras, MOS-1 and MOS-2   \citep{turner2001}, 
were set to full window (FW) and TM, respectively.

We ran the {\tt XMM-SAS} pipeline analysis scripts (SAS-version 18.0) for all EPIC and RGS instruments on board \xmm. We verified that the background flares were not detected, therefore, it was not necessary to perform further cleaning.  We extracted the source spectrum from MOS-2 in TM from the {\tt RAWX} interval  [285, 325], while the background spectrum was extracted from MOS-2 in the image mode  from a circle region with a radius of $150\arcsec$.  The response matrix file  and  ancillary response file were generated using {\tt rmfgen} and {\tt arfgen}, respectively. 

For the timing analysis, we used data from the \xmm\ EPIC pn. These were subsequently barycentered using the {\tt SAS barycen 1.21} script. 
Furthermore, we selected the one-dimensional spatial parameter {\tt RAWX} by defining the source-region as {\tt RAWX} interval [31,44]  and the background region as the combination of {\tt RAWX} [11,19] and [55,63], chosen far from the source region. The latter has been used in the estimation of background corrected fractional amplitudes in the pulse-profile analysis.

The (non-imaging) Reflection Grating Spectrometers (RGS) \citep{denherder2001}, on board \xmm,\ operated in default mode (HighEventRate with Single Event Selection; HER + SES), collecting 
spectral information in the 0.35--2.5~keV band. Two RGS spectra  were extracted using {\tt rgsproc} and the corresponding response files were created using {\tt rgsrmfgen}. In order to increase the signal-to-noise ratio (S/N), we combined the spectra of two RGS data in first order.  
\xmm\ did not observe \psrtar\ during its 2019 outburst.

%We also checked the data for the presence of soft proton background flares, but we detected none, and therefore further cleaning was not required. 

\subsection{\hxmt}
\label{sec:hxmt}

The first Chinese X-ray telescope, the Hard X-ray Modulation Telescope (HXMT), was launched on June 15, 2017  and later dubbed \hxmt\ \citep{hxmt}.
Three slat-collimated instruments, the Low Energy X-ray telescope \citep[LE, 1--15~keV, 384~cm$^2$;][]{hxmt-le}, the Medium Energy X-ray telescope \citep[ME, 5--30~keV, 952~cm$^2$;][]{hxmt-me} and the High Energy X-ray telescope \citep[HE, 20--250~keV, 5100~cm$^2$;][]{hxmt-he} on board \hxmt\ provide the capability for the broadband X-ray timing and spectroscopy. \hxmt\ observed \psrtar\ during its 2018 outburst on MJD 58213.1--58213.8.
We employed the HE and ME data to investigate the timing properties of the source because
of their relatively good time resolution ( $\sim$2~$\mu$s for HE, $\sim$20~$\mu$s for ME).  We analyzed  the data using the \hxmt\
Data Analysis Software package (HXMTDAS) version 2.01. The ME and HE data were calibrated  by using the scripts  {\tt mepical} and {\tt hepical}, respectively. The good time intervals were individually selected from the scripts {\tt megtigen} and {\tt hegtigen} with the loose criteria, that is, ELV > 0 and the satellite located outside the SAA region. 
Finally, the events were obtained using {\tt mescreen} and {\tt hescreen} and  were barycentered by the tool {\tt hxbary}. \hxmt\ did not observe \psrtar\ during its 2019 outburst.

\subsection{\swift}
\label{sec:swift}

In total, nine and one observations, respectively, are available for the 2018 and 2019 outbursts from  \swift/XRT, respectively. The \swift/XRT light curves were only reduced to construct the outburst profiles during the 2018 and 2019 outbursts. We reduced the \swift/XRT data in the photon counting mode. The pipeline {\tt xrtpipeline} was operated for each observation and  the light curve was extracted from a circle region centered on the source position with a radius of $25\arcsec$ and  corrected by the  \swift\ tool {\tt xrtlccorr}.

%%%%%%%%%%%%%%%%%%%%%%%% Fig-3 %%%%%%%%%%%%%%%%%%%%%%%%
\begin{figure} %[t]
\centering
\includegraphics[width=8.5cm]{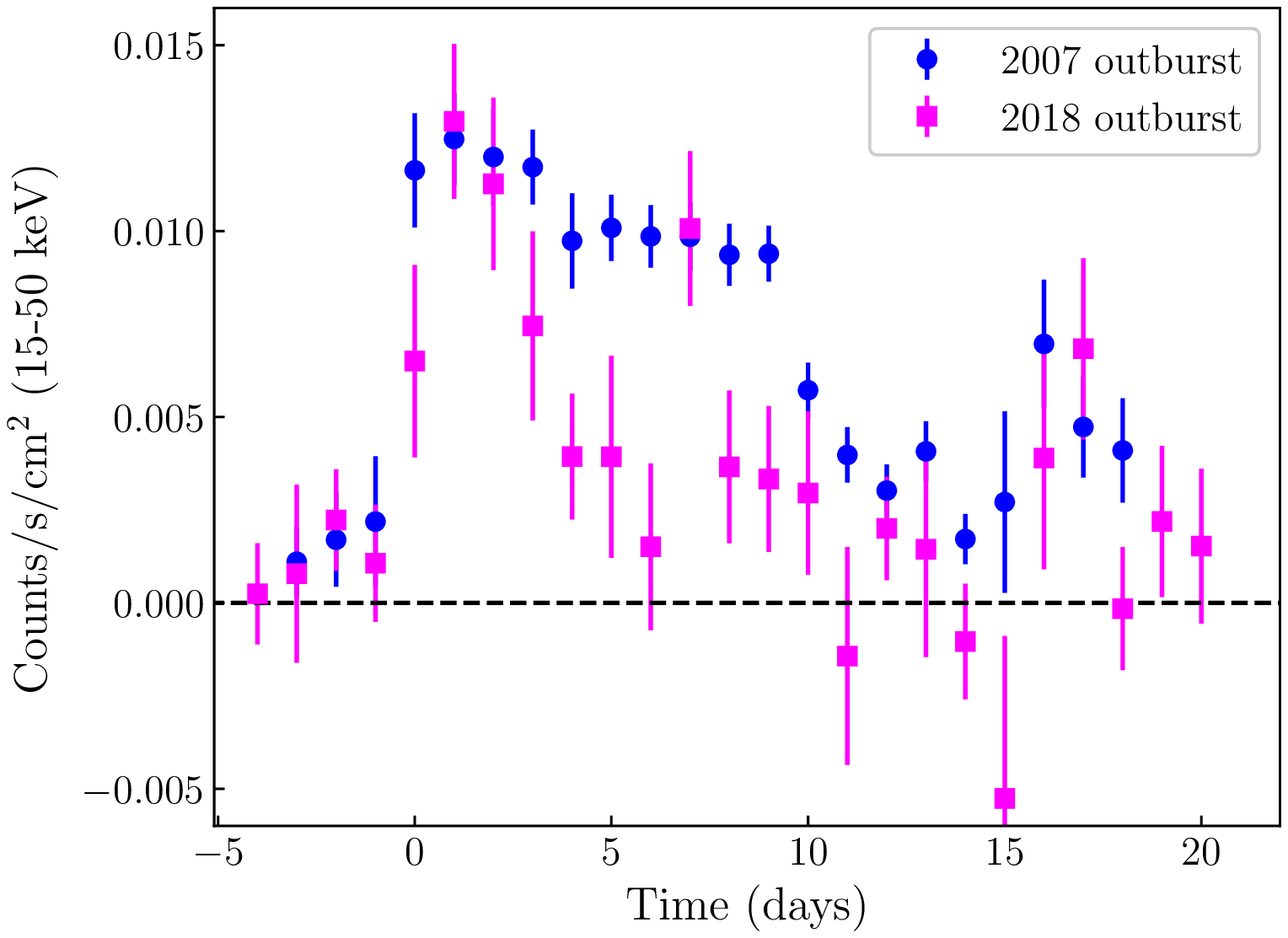}
\includegraphics[width=8.5cm]{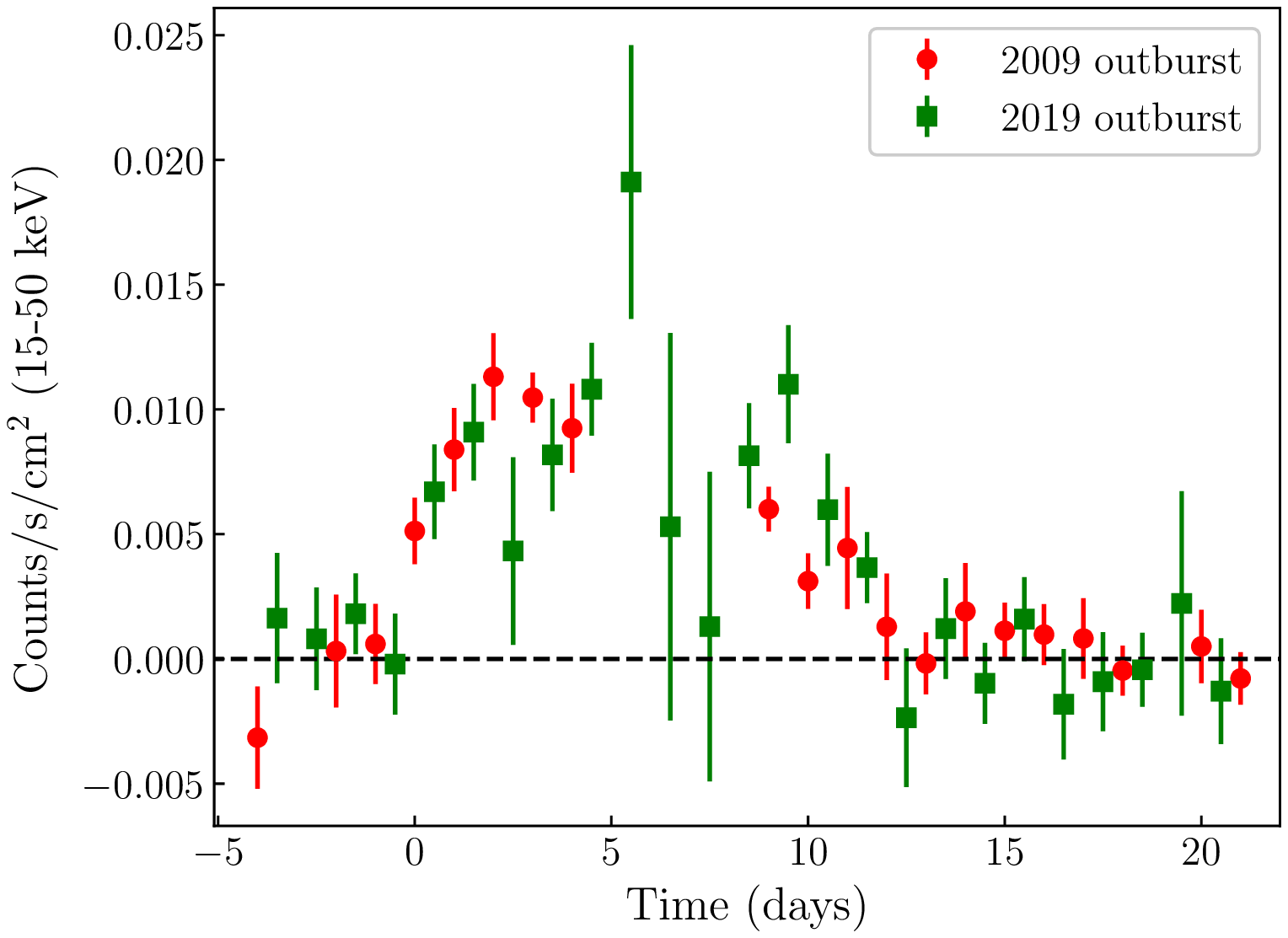}

\caption{{\it Swift}-BAT daily averaged (15--50 keV) profiles of the 2007 (blue dots) and 2018 (magenta squares) outbursts (top panel), the 2009 (red dots), and 2019 (greed squares) outbursts (bottom panel) from \psrtar. The start time of the 2007, 2009, 2018, and 2019 outbursts are MJD 54258, 55012, 58208, and 58644.5, respectively.
}
\label{fig:bat_outburst}
\end{figure}
%%%%%%%%%%%%%%%%%%%%%%%% Fig-3 %%%%%%%%%%%%%%%%%%%%%%%%

%%%%%%%%%%%%%%%%%%%%%%%% Spectrum %%%%%%%%%%%%%%%%%%%%%%%%
\begin{figure*} %[h]
%\begin{center}
\hbox{
\includegraphics[angle=-90,width=8.5cm]{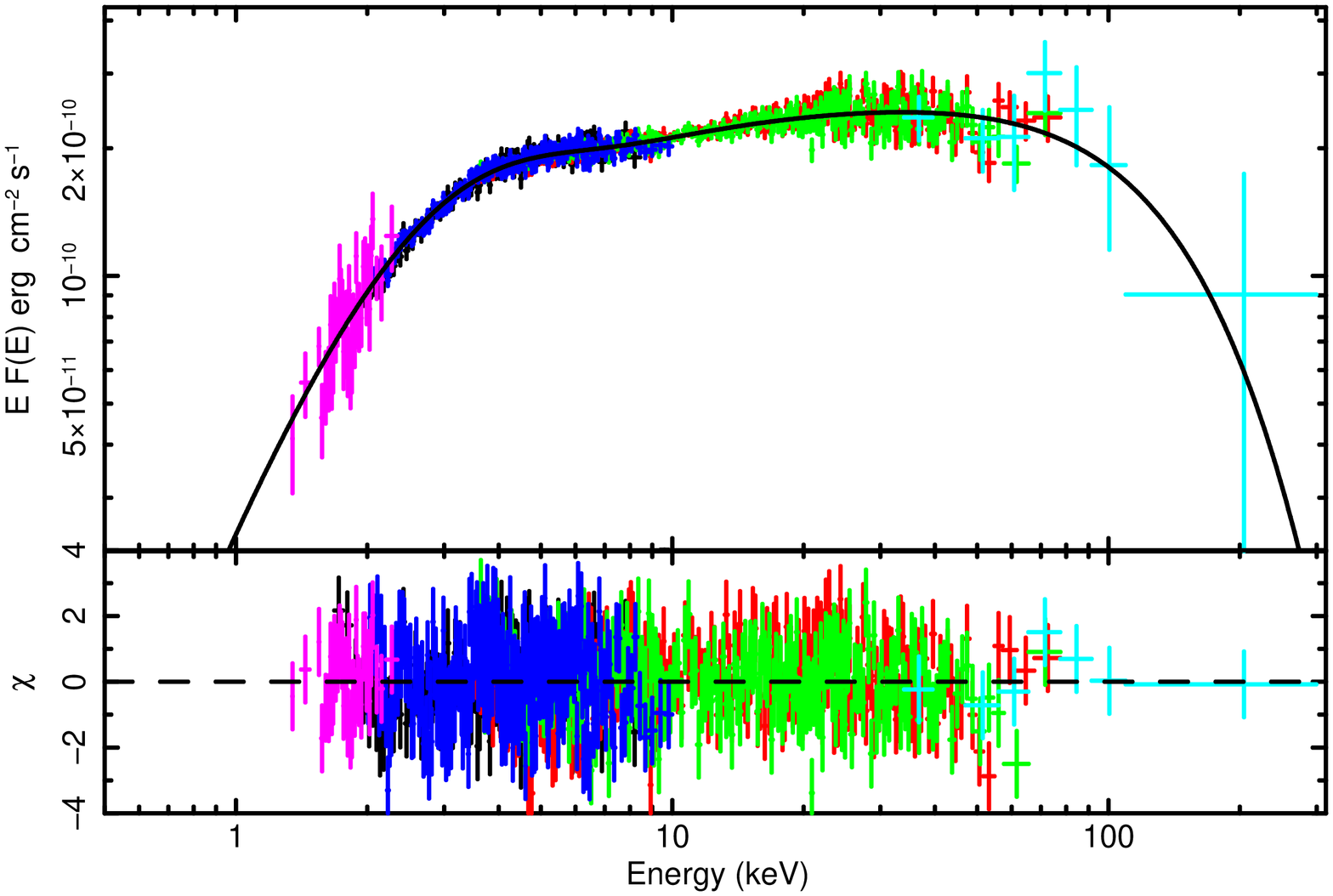}
\hspace{+0.5cm}
\includegraphics[angle=-90,width=8.5cm]{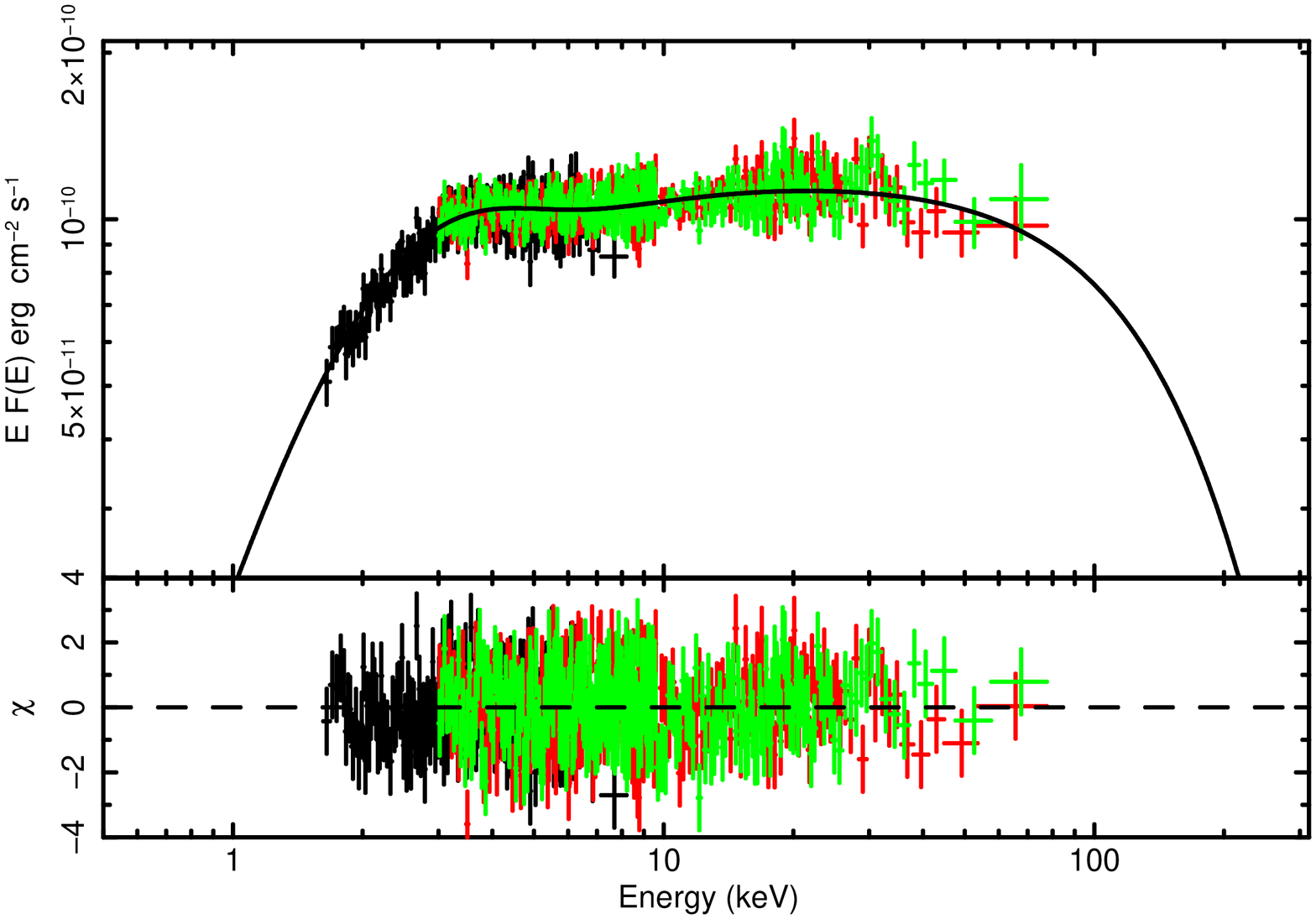}
}
%\vspace{0.2cm}
%\end{center}
\caption{
%\footnotesize{
Unfolded and unabsorbed broadband spectra of \psrtar.
{\it Left panel:} Spectrum for the 2018 outburst in the 1--250 keV energy range. 
The data points are obtained from the combined of two \xmm\ RGS instruments (magenta  datapoints, $\sim$1.2--2.4 keV), \xmm\ EPIC-MOS2 (blue points, 2--10 keV), \nicer\ (black points, 1.5--10 keV) \nustar\ FPMA/FPMB (red and green  points, 3.5--80 keV), and \Integ-ISGRI (cyan points, 30--250 keV). {\it Right panel:} Spectrum of the 2019 outburst in the 1.5--80 keV energy range. The data points are obtained from the \nicer\ (black  points, $\sim$1.5--10 keV), \nustar\ FPMA/FPMB (red and green  points, 3--80 keV). In both cases, the fits are obtained with the \compps\ model, represented with a solid line. The residuals from the best fit are shown at the bottom. 
%}
}
\label{fig:spec}
\end{figure*}
%=======================

\section{The outburst profiles}
\label{sec:outburst_profile}

The light curves of \psrtar,\ as obtained from all available X-ray data, show the entire 2018 and 2019 outbursts both lasting  about 15~d (from March 31 to April 14, 2018 and June 10 to 25, 2019) are shown in Fig.~\ref{fig:outburst_profile}. Because the 2019 outburst was observed at the last stage during its decay phase, the flux in one-day bins was estimated using the \swift-BAT data. The count-rates measured from all instruments were converted to bolometric flux (0.1--300 keV) using the spectral analysis results obtained in Sect.~\ref{sec:spectrum} for the respective outbursts.  

% new para here --- dkg
The profile of the outburst observed from \psrtar\ is not too dissimilar from those shown by other AMXPs, which are typically characterized by a swift rise time ($\sim$2--3~d), followed by $\sim$7 d exponential and $\sim$2~d linear decay down to the quiescence level  \citep[see e.g.,][]{mfb05,falanga11,falanga12,ferrigno2011,bozzo16,deFalcob,deFalcoa,Kuiper20}. Using the best covered 2018 outburst profile, we estimate the outburst peak flux around $\sim 2.0 \times 10^{-9}~ {\rm erg~cm^{-2}~s^{-1}}$ in the 0.1--300 keV energy range (see Fig. \ref{fig:outburst_profile}); this value is similar to that for all other observed outbursts, considering the same energy band \citep{Krimm07a,Linares08,Patruno09,Sanna18}.

\begin{table} %[h] 
{\small
\caption{Best parameters determined from the fits to the  broadband emission spectra 
of \psrtar\ using the {\sc phabs} absorbed comptonization model \compps. In the 2018 outburst, the spectra were obtained from  the \xmm, \nicer, \nustar,\ and \Integ. The  \nicer\ and \nustar\  spectra were utilized for the 2019 outburst. }
\centering
\begin{tabular}{lccc} 
\hline 
Parameter & Unit & 2018 outburst & 2019 outburst \\
\hline 
\noalign{\smallskip}  
$N_{\rm H}$                  & ($10^{22}$~cm$^{-2}$) & $4.29\pm0.05$ & $4.17\pm0.12$\\ 
$kT_{\rm bb, ~seed}$          &(keV)                      & $0.79\pm0.01$ & $0.67\pm0.02$\\ 
$kT_{\rm e}$                 &(keV)                      & $41.6\pm0.7$  & $44.6\pm1.3  $\\ 
$\tau_{\rm T}$               &                           & $1.44\pm0.02 $   & $1.23\pm0.03$\\ 
%$A_{\rm bb}$\tablefootmark{a}              &(km$^2$)      &$ 96\pm6    $       & $100\pm14 $\\
$K_{\rm bb}$              &(km$^2$)      &$ 150\pm9    $       & $156\pm20 $\\
$\chi^{2}_{\rm red}/{\rm d.o.f.}$  &  & 1.11/1644      & 1.01/858 \\
$F_{\rm bol}$\tablefootmark{a}    & ($10^{-10}$~erg cm$^{-2}$ s$^{-1}$) & $8.63\pm0.06$ & $4.44\pm0.05$\\
\noalign{\smallskip}  
\hline  
\end{tabular}  
\tablefoot{
Uncertainties are given at $1\sigma$ confidence level.
%\tablefoottext{a}{The apparent area was estimated assuming the distance $d=8$~kpc. }
\tablefoottext{a}{Unabsorbed flux in the 0.1--300 keV energy range.}
}
\label{table:spec} 
}
\end{table} 
%=======================

Such outbursts are typically described by the disk instability model  \citep[see][for more details]{King1998}, which predicts that both linear and exponentially decaying outbursts can be produced. 
%When the irradiation of the accretion disk from the X-ray emission of the NS is strong enough (i.e., the peak X-ray luminosity at the onset of the outburst is high), the disk is completely ionized out to its outer edge and the light curve of the outburst follows an exponential decay. In this phase, most of the disc mass can be accreted onto the NS in a viscous timescale. A change in the profile of the outburst is expected once the source has faded below a certain luminosity level. Below this threshold, only part of the accretion disk can be ionized. As a consequence, both the mass accretion rate onto the NS and the outer boundary of the ionized region in the disk begin to decrease with time, leading to a linear (rather than exponential) decay.
% These outburst profiles can be well described within the disk instability model, i.e. 
Typically, for such outbursts, the flux decays from the maximum exponentially until it reaches a break (``knee"), in our case around MJD~58217 (2018 outburst) or MJD~58656 (2019 outburst), and then linearly drops to the quiescent level.  Following \citet{Powell2007}, the outer disk radius can be estimated by fitting the decay profile (see Fig. \ref{fig:outburst_profile}) with the expression:
\begin{equation} 
L_{\rm X} = (L_{\rm t}-L_{\rm e})\, \exp[ -(t-t_{\rm t})/\tau_{\rm e}] + L_{\rm e}, 
\end{equation} 
where $L_{\rm e}$ (the limit luminosity of the exponential
decay),  $t_{\rm t}$ (the break time), $L_{\rm t}$ (the luminosity at the break time $t_{\rm t}$), and $\tau_{\rm e}$ (exponential decay time) are all free parameters. The outer disk radius is $R_{\rm disk}(\tau_{\rm e})= (\tau_{\rm e}\,3\nu_{\rm KR})^{1/2} \sim
2.5\times10^{10}$ cm, where  the value of the viscosity near the outer disk edge  $\nu_{\rm KR}= 10^{15}$ cm$^2$ s$^{-1}$ is adopted \citep[see][ for more details]{King1998,Powell2007}. In the 2019 outburst, we obtain the outer disk radius of $R_{\rm disk}\sim
3.0\times10^{10}$ cm. 
For the binary inclination of $60^\circ$ and the NS mass of $1.4M_\odot$, the companion mass $M_{\rm c}$ is $0.0078M_\odot$. We adopted the relation 
\begin{equation}
    \frac{b_1}{a}=0.741-1.682q+12.053q^2-42.342q^3 -0.0663\log_{10}q,
\end{equation}
where $b_1$ is the distance between the NS and the Lagrange point, $L_1$, of the binary system and $a$ is the binary separation, which has an accuracy of 0.5\% for
$q$ between 0.0025 and 0.11 \citep{Iaria21}. The circularization radius $R_{\rm circ}$ can be estimated from \citep[see e.g.,][]{frank02}:
\begin{equation}
   \frac{R_{\rm circ}}{a}=(1+q)(b_1/a)^4.
\end{equation}
We found that the outer disk radii in the 2018 and 2019 outbursts satisfy the condition $R_{\rm circ} < R_{\rm disk} < b_{1}$, where $R_{\rm circ} \approx 2.3 \times 10^{10}$ cm  and $b_{1} \approx 3.3 \times10^{10}$ cm.

In Fig.~\ref{fig:bat_outburst}, an inspection of all the \swift-BAT outburst data shows that the period of activity in 2018--19  broadly follows the pattern of the previous outbursts, with two more closely-spaced outbursts separated by a longer interval. The 2009 outburst followed the discovery outburst in 2007 by 2.1~yr; the 2019 outburst came 1.2~yr after 2018. However, the 2018 outburst came after a much longer quiescent interval, of 8.7~yr. 

%that the profiles of the 2007 and 2018 outbursts are almost identical with a rise time of $\sim2$  days, as are the 2009 and 2019 outbursts with a slower rise time of $4\sim5$ days,  and these outbursts happen at irregular recurrence times. T

Such pairs of outbursts have only previously been observed from the low-mass X-ray binaries IGR J00291+5934 and XTE J1118+480 \cite[]{hartman11}, although in those systems, the inter-outburst separation was 30~d. Those authors attributed the secondary outburst to leftover material in the accretion disk that was not deposited on the NS during the first outburst. The atypical mass distribution left in the disk when the accretion stalls during the first outburst, leads to the unusual slow-rise shape of the second outburst; this pattern is very similar to what is observed in \psrtar.

Similarly, for those sources, the wide variation in outburst separation in \psrtar\/ is not consistent with a steady accretion rate into the disk, unless the bulk of the fuel for the second outburst in each pair was leftover material that was not accreted during the first outburst. We estimate here the fraction of fuel leftover, and the lower limit on the steady accretion rate, as follows. By integrating the flux measurements over each outburst, the fluences for the 2018 and 2019 outbursts are $1.1\times10^{-3}$~erg~cm$^{-2}$ and $1.2\times10^{-3}$~erg~cm$^{-2}$, respectively. The fluence of each outburst arises from an accreted mass of: 
%\begin{eqnarray}
\begin{equation}
\Delta M  =  \frac{4\pi d^2 \int F_{\rm X} dt}{Q_{\rm grav}} 
\approx  \frac{4\pi d^2 R_{\rm NS} \int F_{\rm X}dt}{GM_{\rm NS}}
,\end{equation}
%\end{eqnarray}
where $Q_{\rm grav}\approx GM_{\rm NS}/R_{\rm NS}$ is the gravitational potential energy liberated during accretion. For each of the 2018 and 2019 outbursts, the accreted mass is $\Delta M \approx 3\times10^{-11}\ M_\odot d_8^2 r_{10} m_{1.4}^{-1}$, where $d_8$ is the distance in units of 8~kpc, $r_{10}$ the NS radius in units of 10~km, and $m_{1.4}$ the NS mass in units of $1.4M_\odot$. Assuming a constant accretion rate $\dot{M}$ over the interval between the 2009 and 2019 outbursts, we can solve for this quantity as well as the amount of fuel left over following the 2018 outburst. We find a steady accretion rate of $5\times10^{-12}\ M_\odot\,{\rm yr}^{-1} d_8^2 r_{10}\,m_{1.4}^{-1}$, implying about 45\% of the material in the disk remaining after the end of the 2018 outburst. 

% added 13.1.21
% formula from Galloway et al. 2006, adapted in turn from Bildsten & Chakrabarty (2001);
% sqrt(3.8e-11/5e-12*(54.7/60./2)**(-8./3.)*(6.7e-3/0.1)**2)*8 = 4.212
We can estimate a lower limit for the distance to the source by equating the implied long-term accretion rate above, with the expected accretion rate driven by gravitational-wave radiation from the binary orbit \cite[e.g.][]{bc01}. Adopting the minimum companion mass of $M_c=6.7\times10^{-3}\ M_\odot$ for a $1.4M_\odot$ NS \cite[]{krimm2007}, we find that $d_8^2 \gtrsim 0.277\, m_{1.4}^{5/3}\,r_{10}^{-1}$, so that $d\gtrsim4$~kpc. This limit is consistent with our assumed distance of 8~kpc.

\section{Broadband spectral analysis of the 2018 and 2019 outbursts} 
\label{sec:spectrum}

We studied  the broadband X-ray spectra of the 2018 and 2019 outbursts individually. For the 2018 outburst, we fit the quasi-simultaneous spectra, including the INTEGRAL-ISGRI (20-250 keV) ToO data between MJD 58215.8--58216.8,  the  \nustar\ FPMA/FPMB (3.5--79 keV) data starting on MJD 58216.4  (obs. ID 90402313002), \nicer\ (1.5--10 keV) data starting on MJD 58216.0 (obs. ID 1050230104), \xmm\ MOS (2--10 keV), and RGS (1--2.4 keV) data starting on MJD 58216.1. We note that the MOS spectrum showed a strong excess below 2 keV, which may be affected by the straylight. 
For the 2019 outburst, we consider the simultaneous observations from \nicer\ (1.5--10 keV, obs. ID 2050230104) and \nustar\ FPMA/FPMB   (3--79 keV, obs. ID 00033646004), respectively.  All spectra are grouped to make sure that each channel has more than 25 photons. For each instrument, a multiplication factor is included in the fit to account for the uncertainty in the cross-calibration of the instruments. For all fits, the factor is fixed at 1 for the \nustar\ FPMA instrument. All uncertainties in the spectral parameters are given at a $1\sigma$ confidence level for a single parameter. The spectral analysis were carried out using Xspec version 12.10 \citep{arnaud96}.

We fit all the spectra by using the thermal Comptonization model, \compps,  in the slab geometry \citep{Poutanen96}, with the interstellar absorption described by model {\tt phabs}. This model has been used previously to fit  the broadband spectra of AMXPs \citep[see e.g.,][and references therein]{deFalcoa,deFalcob,ZLI2018,Kuiper20}. The main parameters are the Thomson optical depth across the slab, $\tau_{\rm T}$, the electron temperature, $kT_{\rm e}$, the temperature of the soft seed photons (assumed to be injected from the bottom of the slab), $kT_{\rm bb,~seed}$, the 
%apparent area of the seed photons, $A_{\rm bb}$, 
normalization factor for the seed blackbody photons, $K_{\rm bb}=(R_{\rm km}/d_{10})^2$ (with $d_{10}$ being distance in units of 10 kpc), 
and the inclination angle $\theta$ (fixed at 60\degr) between the slab normal and the line of sight to the observer.
The interstellar absorption is described by the  hydrogen column density, $N_{\rm H}$.
The best-fit parameters for all models are reported in Table \ref{table:spec}.  
For the distance of 8 kpc, the size of the blackbody emitting region $R_{\rm bb}=0.8\sqrt{K_{\rm bb}}\approx 10$~km.
This is very similar to IGR~J17498$-$2921 \citep{falanga12}, which was also assumed to be situated at the same distance, but much larger, for instance, than in XTE~J1751$-$305 \citep{gp05}. 
Such a large size is hardly consistent with the hotspot at the NS surface and might indicate that the distance to the source is smaller. 
The bolometric fluxes are calculated by the convolution model {\tt cflux} in the 0.1--300 keV energy range. In Fig. ~\ref{fig:spec}, we show the best-fit spectra.

\begin{figure*}[t]
  \centering
  \includegraphics[width=16cm]{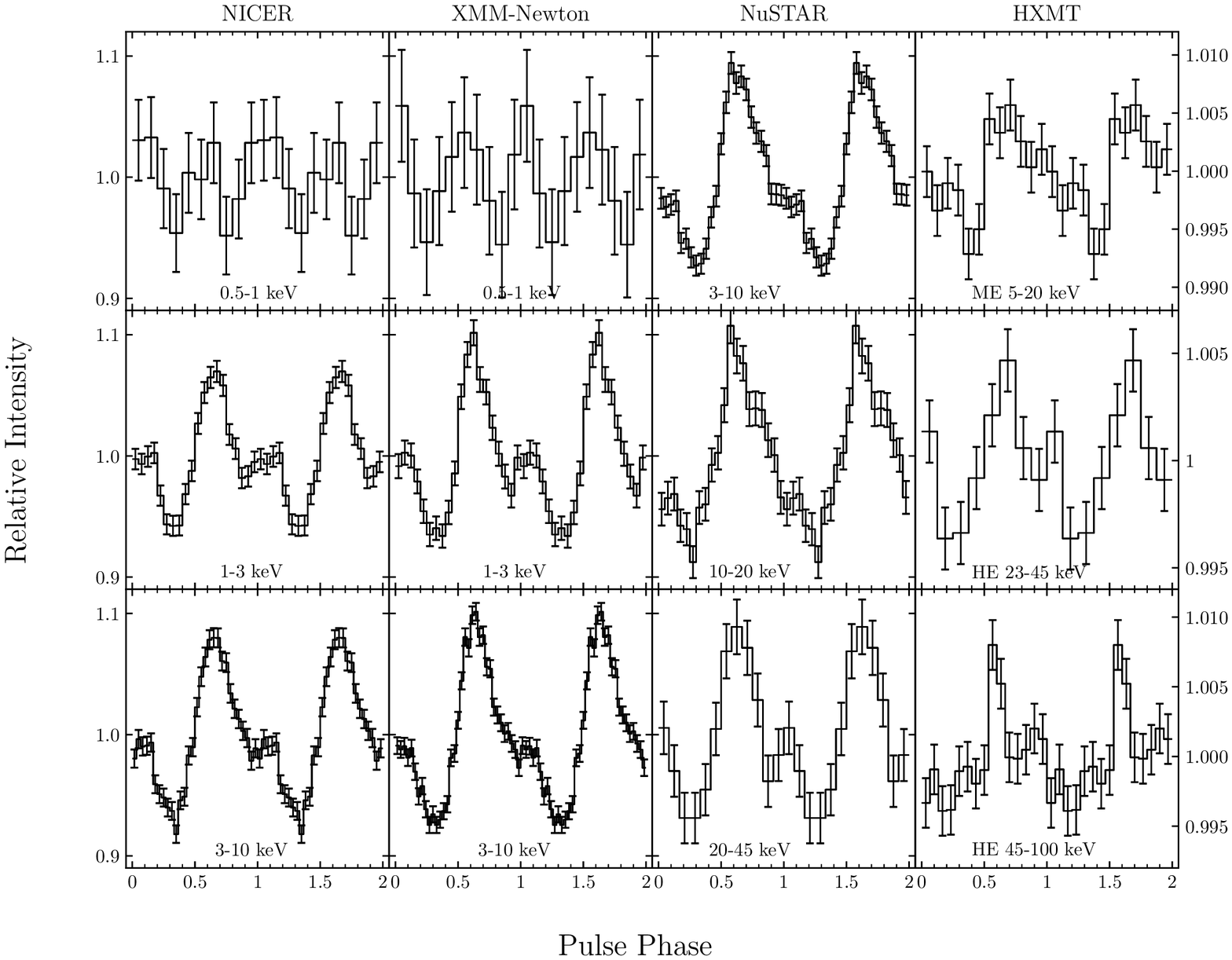}
  \caption{Pulse profiles of \psrtar\ observed by \nicer, \xmm, \nustar, and \hxmt\ during the 2018 outburst. }
  \label{profile_2018}
\end{figure*}

We find that all multiplication cross-calibration factors are around unity, which means the flux calibration of these instruments is well established and the source did not vary much between the quasi-simultaneous observations. The 2018 and 2019 outbursts can be characterized by similar hard spectra (see Table~\ref{table:spec}). This indicates that even thought the 2019 outburst was observed at a later outburst stage, namely, at a flux level lower by a factor 1.9 , the decay of the outburst is characterized by a nearly constant spectral shape described by similar electron temperature, $kT_{\rm e}$, soft seed photons temperature, $kT_{\rm bb,~seed}$, and the optical depth, $\tau_{\rm T}$. The spectral parameters are similar to those of many other AMXPs \citep[see e.g.,][and references therein]{falanga11}. 
The spectra for the 2018 and 2019 outbursts in narrower energy ranges are well described by a power-law with a photon index in the range of $\sim$1.8--2.0, which is compatible to those measured in the 2007 and 2009 outbursts  \citep{Krimm07a,Linares08,Patruno09}. 
Also, the hydrogen column density was similarly high, namely, $N_{\rm H}\sim(4.5-5.4)\times10^{22}~\rm{cm^{-2}}$. 
This also confirms that all four observed outbursts are very similar (see also Sect. \ref{sec:outburst_profile}). 
We note that our estimate of the blackbody normalization
%of the area of the blackbody emission region $A_{\rm bb}$
is at least factor of 30 larger than the normalization quoted by \citet{Sanna18}, who used a phenomenological blackbody plus cutoff power-law model. 
The difference comes from the facts that in the \compps\ model, only a small fraction, that is, $\exp(-\tau_{\rm T}/\cos\theta)\sim 0.06$, of the seed blackbody photons pass through the Comptonizing slab unaffected; the rest is scattered and produce the hard tail extending to 100 keV. 
Also, in our model, a large fraction of the flux below 3 keV comes from the blackbody, whereaws in the model of \citet{Sanna18}, the power law contributes most of the flux.

\begin{table}[t] 
{\small
\caption{Positional, rotational, and orbital parameters used from other works and derived in this work for \psrtar.}
\centering
\begin{tabular}{lcc} 
\hline 
Parameter &Units & Values  \\
\hline 
\noalign{\smallskip}  
$\alpha_{2000}$                &                              & $17^{\hbox{\scriptsize h}} 56^{\hbox{\scriptsize m}} 57\fs350$  \\   
$\delta_{2000}$                &                              & $-25\degr06\arcmin27\farcs80$ \\ 
JPL Ephemeris                  &                              & DE405\\
$ P_{\rm orb} $                &s                             & $3282.3515$ \\
$ a_{\rm x}\, \sin i$          &lt-s                          & $0.00597$ \\ 
$ e $                          &                              & $0.00$ \\
\vspace{-0.25cm}\\
\multicolumn{3}{c}{Outburst - 2018}\\
\vspace{-0.25cm}\\
$\nu$                          &Hz                            & $182.065\,803\,84(3)$\\
Epoch, $t_0$                   &MJD; TDB                      & $58216.0$\\
Validity range                 &MJD; TDB                      & $58211 - 58219$  \\
$T_{\rm asc, 2018} $           &MJD; TDB                      & $58211.017\,52(6)$ \\
\vspace{-0.25cm}\\
\multicolumn{3}{c}{Outburst - 2019}\\
\vspace{-0.25cm}\\
$\nu$                          &Hz                            & $182.065\,803\,4(2)$\\
Epoch, $t_0$                   &MJD; TDB                      & $58654.0$\\
Validity range                 &MJD; TDB                      & $58653 - 58657$  \\
$T_{\rm asc, 2019} $           &MJD; TDB                      & $58655.996\,57(12)$ \\
%$t_0$                         &MJD; TDB                      & 58439  \\
%$\nu$                         &Hz                            & 182.0658038016(4)\\
\noalign{\smallskip}  
\hline  
\end{tabular}

\label{table:eph} 
}
\end{table} 

\section{Timing analysis}
\label{sec:timing}

Irrespective of the instrument, in timing analyses, we had to convert the Terrestial Time (TT) arrival times of the (selected) events to arrival times at the solar system barycenter (in a TDB time scale). For  this process, throughout in this work we used: 1) the JPL DE405 solar system ephemeris; 2) the instantaneous spacecraft position with respect to the Earth's center; and 3) the X-ray celestial position of \psrtar, $\alpha_{\rm J2000} = 17^{\rm h}56^{\rm
m}57\fs35$ and $\delta_{\rm J2000} = -25{\degr}06\arcmin27\farcs8$ obtained by the \swift-XRT telescope \citep{krimm2007}.

\subsection{\nicer\ timing analysis of the 2018 and 2019 outbursts}

For our timing analysis, we selected "cleaned"  barycentered NICER XTI events, collected during the 2018 and 2019 outbursts, from the standard pipeline analysis with measured energies between 0.5 and 10 keV. Events with energies between 12–15 keV, however, were used to flag periods with high-background levels (e.g., South Atlantic Anomaly ingress or egress, etc.) as bad, and these intervals have been ignored in further analyses. Moreover, events from noisy or malfunctioning detectors were ignored. The screened events were subsequently barycentered using a (multi-instrument serving) {\sc idl} procedure.

\begin{figure}[t]
  \centering
  \includegraphics[width=9.8cm]{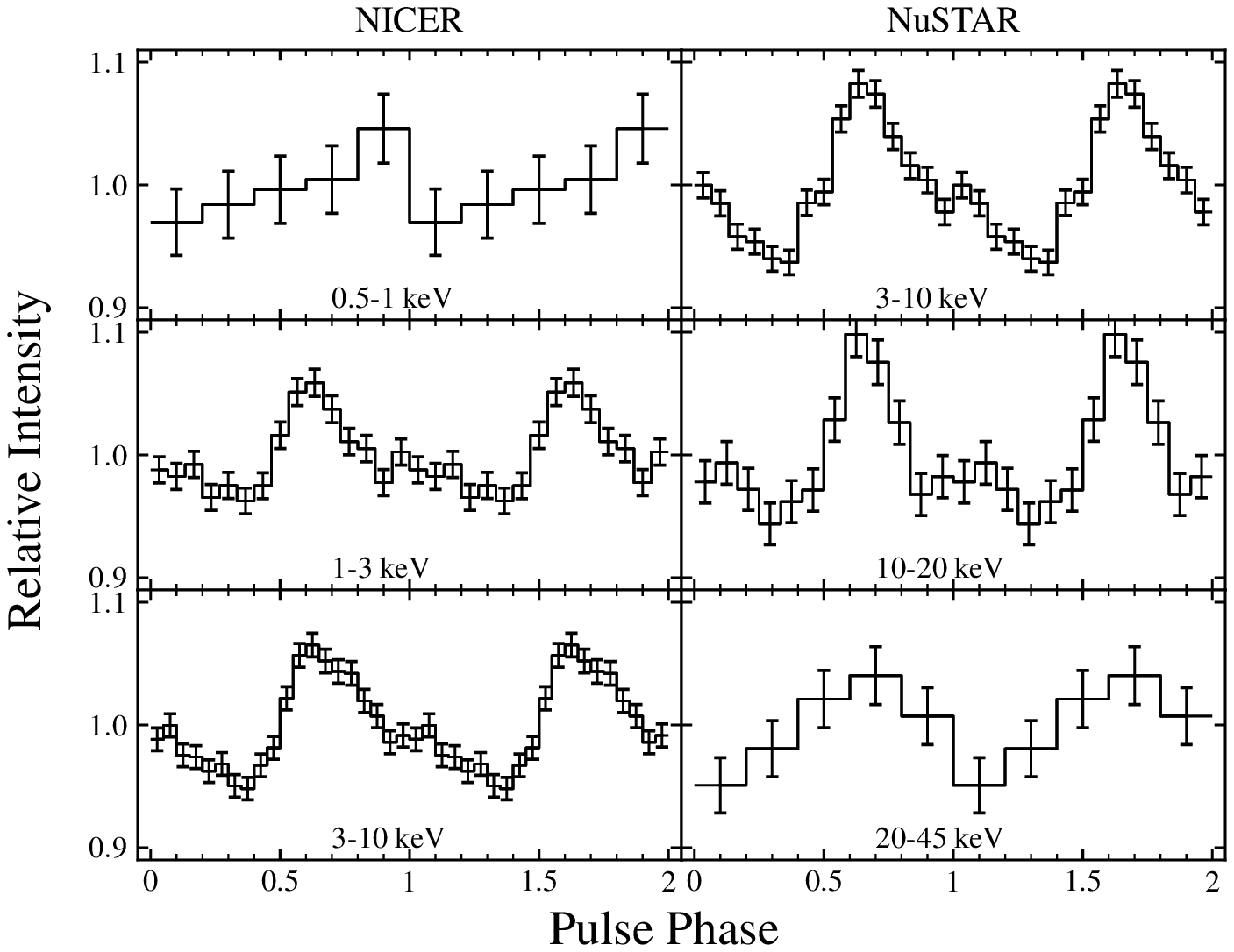}
  \caption{Pulse profiles of \psrtar\ observed by \nicer\ and \nustar\ during the 2019 outburst. }
  \label{profile_2019}
\end{figure}

For each outburst,  we individually determined the pulse frequency, $\nu,$ and time of ascending node, $T_{\rm asc}$, in a 2d-optimization scheme based on a SIMPLEX algorithm \citep[see, e.g.,][for more details]{deFalcoa, Kuiper20}, finding the global maximum of the $Z^2(\phi)$-test statistics \citep{buccheri1983}. In this approach we kept the orbital period, $P_{\rm orb}$, projected semi major axis, $a_x\sin i$, fixed at the optimum values given by \citet{Patruno10} (see their Table 4). We verified that using updated orbital ephemeris information \citep{Bult18} had no impact on the $T_{\rm asc}$ values derived here. The best fit values for  $\nu$ and $T_{\rm asc}$,  along with the validity interval and epoch $t_0$ for both outbursts,  are individually listed in Table \ref{table:eph} and are used later in this work in estimating the orbital period derivative combining information from all registered outbursts (see Sect. \ref{sect_orb}).
Equipped with  accurate timing models for both outbursts (see Table \ref{table:eph}), we phase-folded the barycentered event times from different instruments for different energy bands both for the 2018 and 2019 outbursts. The results are shown in Figs. \ref{profile_2018} and \ref{profile_2019} for the 2018 and 2019 outbursts, respectively.

During the 2018 (also for the 2019) outburst, no pulsed emission was detected below 1 keV using \nicer\ and \xmm\ data. \nustar\  detected pulsed emission up to $\sim 60$ keV during the 2018 outburst. Folding \Integ-ISGRI  barycentered data of the 2018 outburst  for the 20-60 keV band did not result in a detection of the pulsed emission in line with the expectations given the moderate total outburst flux, low exposure, and low pulsed fraction of $\la 8\%$ (see Sect. \ref{sec:pulse_time_lag}). However, with the \hxmt\ HE instrument, significant pulsed emission has been detected up to $\sim 100$ keV (see lower right panel of Fig. \ref{profile_2018}) in spite of relatively low exposure. This suggests great prospects for future observations with \hxmt\ of the AMXP outbursts.

\begin{figure*}[t]
  \centering
  \includegraphics[angle=0,width=8.5cm,height=8cm]{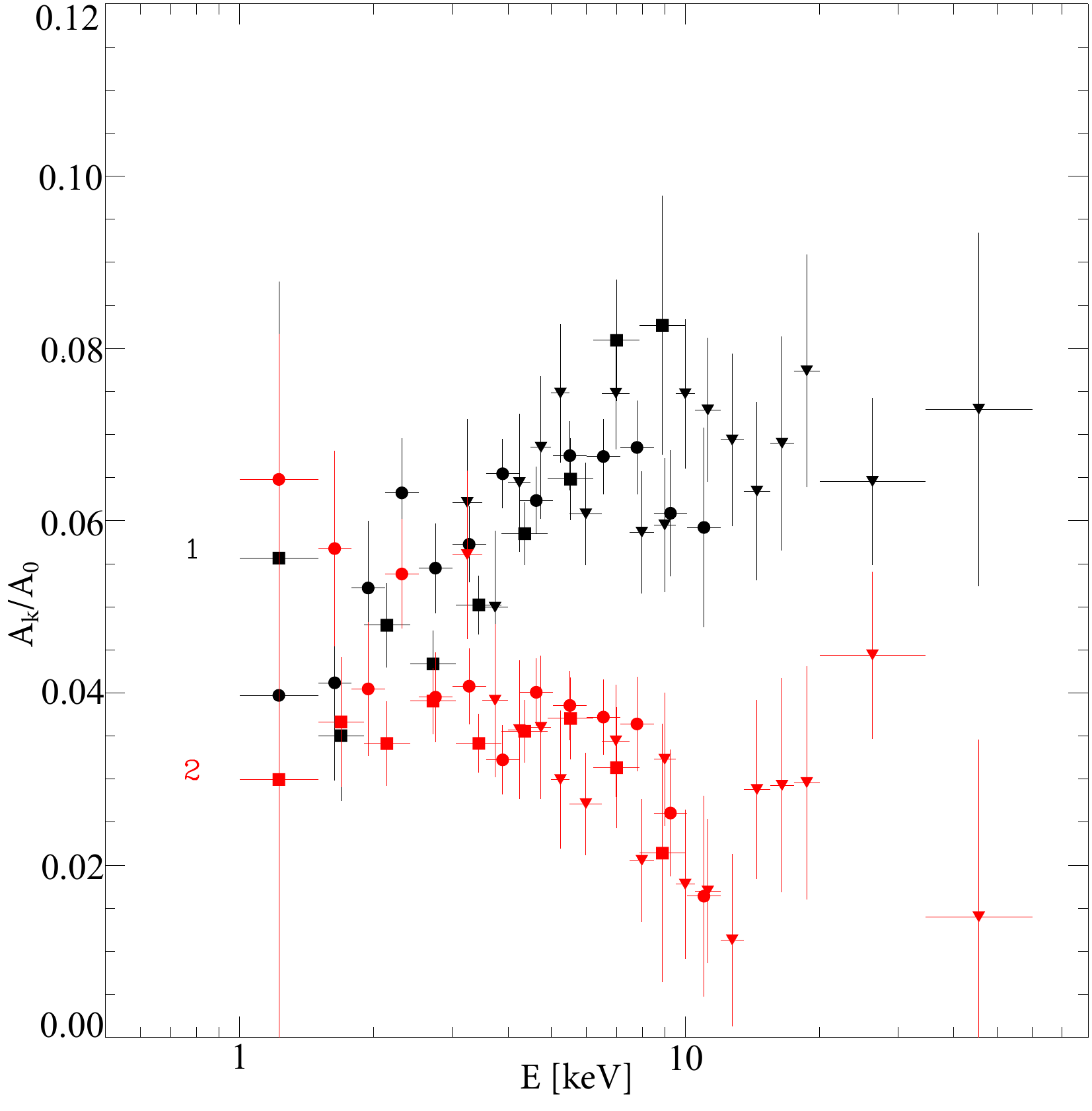}
  \hspace{0.5cm}
  \includegraphics[angle=0,width=8.5cm,height=8cm]{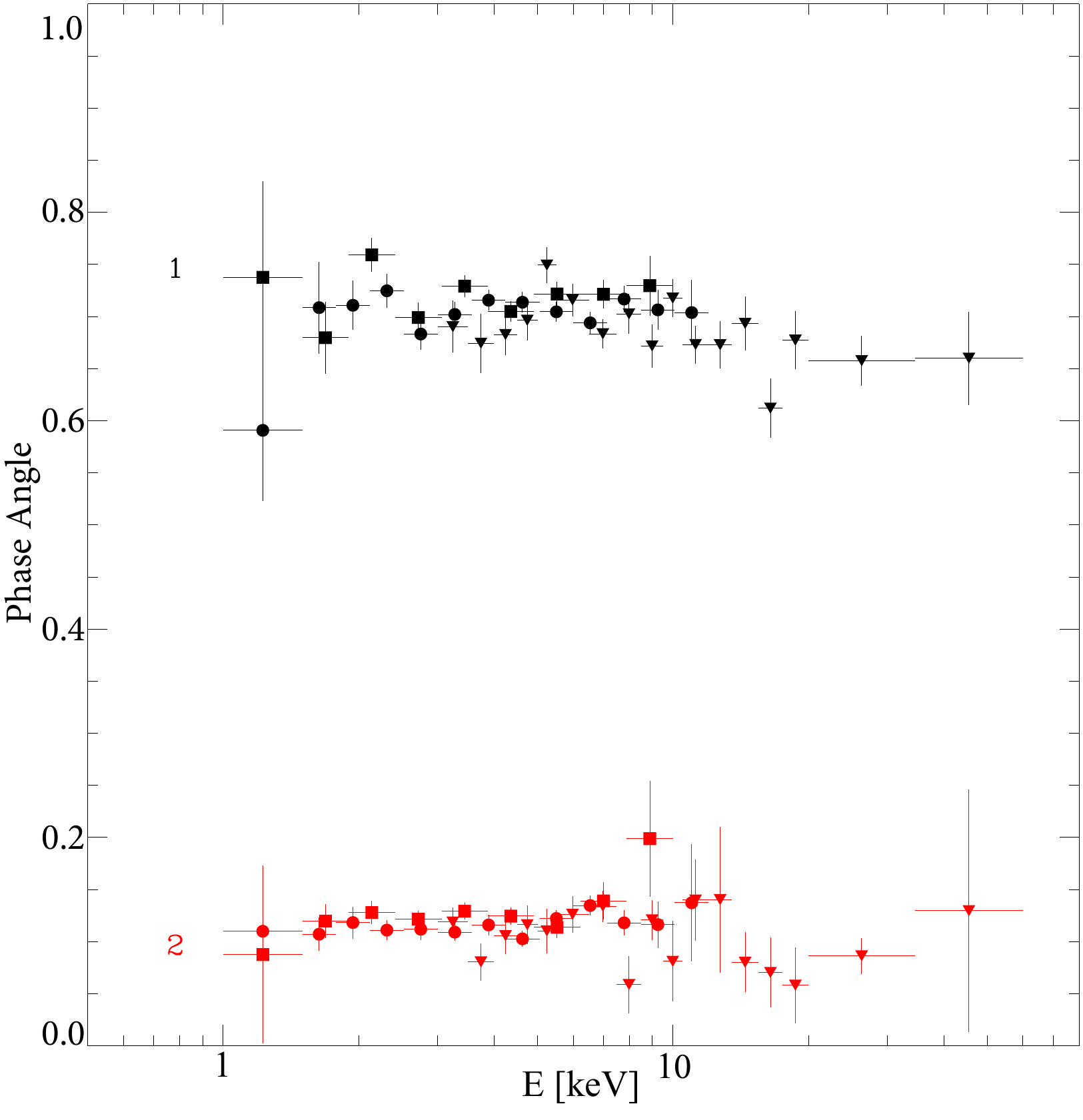}
  \caption{Results of pulsed amplitudes and phase angles. {\it Left panel:}  Fractional amplitude of two harmonics (1 -- fundamental, 2 -- first overtone) as a function of energy using background 
           corrected data from \nicer\ (squares), \xmm\ EPIC-pn (circles) and \nustar\ (triangles).
           The pulsed fraction of the fundamental component increases from $\sim$ 4\% till $\sim 7.5$\% from 1 to 5 keV, from where it more or less saturates at a level of $\sim 7.5$\%.
           {\it Right panel:} Phase angles $\phi_k$ (divided by $2\pi$) for the two Fourier components as function of energy. They remain remarkably stable across the 1--60 keV band, with possibly a small drift for $\phi_1$ to slightly lower values at energies above $\sim$10~keV.}
  \label{pp_fourier_fit}
\end{figure*}

\subsection{Harmonics analysis: Pulsed amplitudes and phase lags}
\label{sec:pulse_time_lag}

We produced pulse-phase distributions in narrow energy bands for \nicer, \xmm,\ and \nustar, covering the $\sim$ 1--60 keV band, to obtain quantitative information about morphology changes of the pulse-profile as a function of energy. 
For this purpose, we fit these measured distributions ${\cal N}(\phi)$ with a truncated Fourier series given by 
\begin{equation}
{\cal F}(\phi)= 
A_0 + \sum_{k=1}^n A_k \,\cos[k\,(\phi-\phi_k)] . \label{eq:fourfit}
\end{equation}
For each harmonic, the maxima occur at $\phi_{\max}=\phi_k \bmod (2\pi/k)$ (in radians).

The results of these fits are shown in Fig. \ref{pp_fourier_fit}. 
The left panel shows the fractional 
amplitude, $A_k/A_0$, and the right panel the phase angle, $\phi_k$, converted from radians to pulse phase, for both the
fundamental ($k=1$, in black) and the first overtone ($k=2$, in red). The different symbols indicated different instruments (\nicer, 1--10 keV, squares; \xmm\ EPIC-pn 1--12 keV, circles, and \nustar, 3--60 keV, triangles).
The pulsed fraction of the fundamental component increases from $\sim$ 4\% up to $\sim 7.5$\% from 1 to 5 keV, from where it more or less saturates at a level of $\sim 7.5$\%. The first overtone fluctuates around 3.5\% with an indication for a depression down to $\sim$ 1\% around 12 keV. These results are consistent with those reported by \citet{Sanna18} (e.g. their Fig. 5). 
Such a behavior was also observed in other sources \citep[see, e.g.,][]{Patruno09b} and is likely related to an increasing contribution of an unpulsed emission from the accretion disk.
The phase angle plot (Fig. \ref{pp_fourier_fit} right panel) shows that the location of both harmonics is stable up to at least 10 keV \citep[see also bottom panel of Fig. 2 of][for equivalent results]{Bult18}, aside from the occurence of a possible small drift to smaller values:
from 0.7 to 0.65 for the fundamental and from 0.12 to 0.07 for the first overtone.

\section{Orbital period }
\label{sect_orb}

\begin{figure}[t]
\centering
\includegraphics[width=9.3cm]{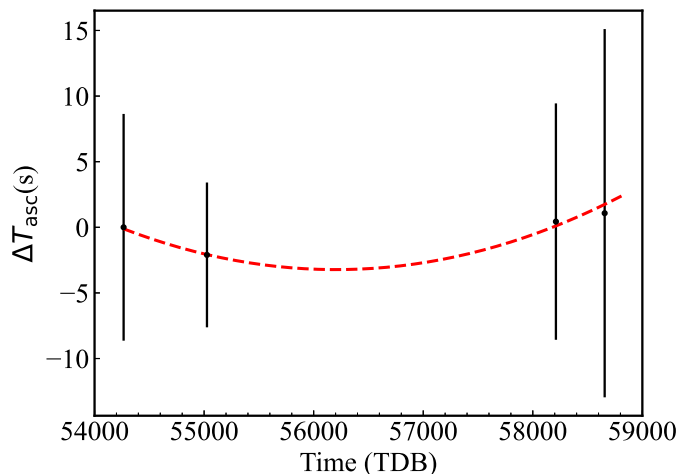}
\caption{Orbital period evolution of \psrtar\ determined from the time of passage at the ascending node during its four outbursts. The red dashed line is the best fit parabolic model.}
\label{fig:orbital_period}
\end{figure}

From the outburst in 2019, we have an additional well-determined ${T_{\rm asc}}$ value. It allows us to perform a coherent analysis of the orbital period evolution across a longer baseline, as compared with \citet{Bult18} and \citet{Sanna18}. Following the procedure introduced in \citet{Hartman2008}, we calculate the residual time of passage through the ascending node: 
\begin{equation}
    \Delta{T_{\rm asc}}=T_{{\rm asc},~ i}-(T_{\rm ref}+NP_{\rm orb}),
\end{equation}
where $T_{{\rm asc},~ i}$ is the time of ascending node determined from the $i$-th outburst, $T_{\rm ref}$ is the reference time, $N$ is the integer number of orbital cycles between the $i$-th outburst and $T_{\rm ref}$, and $P_{\rm orb}$ is the orbital period. We use the orbital period reported in Table~\ref{table:eph} and the reference time $T_{\rm ref}$ as the time of ascending node in the 2007 outburst \citep{krimm2007} to obtain the values of $\Delta{T_{\rm asc}}$ for four outbursts. The errors of $\Delta{T_{\rm asc}}$ are from the uncertainties of $T_{{\rm asc}}$. In Fig.~\ref{fig:orbital_period}, we fit the $\Delta{T_{\rm asc}}$ evolution with a parabolic function, and obtain a best fit value of the orbital period derivative $\dot P$ of $(7.9 \pm 9.5)\times10^{-13}~{\rm s~s^{-1}}$, which is consistent with the results reported in \citet{Sanna18} and \citet{Bult18}. Hence, we conclude that the binary is well described by a constant period over a time span of nearly twelve years.

\section{Non-detection of type-I X-ray bursts}
\label{sect_bursts}

We searched for type-I X-ray bursts in all available \psrtar\ light curves during its 2018 and 2019 outbursts and detected none. The non-detection is consistent with the observations covering the 2007 and 2009 outbursts \citep[][]{Patruno09,Sanna18,Bult18,Rai19}.

%%%%%%%%%%%%%%%%%%%%%%%% Fig-9 %%%%%%%%%%%%%%%%%%%%%%%%
\begin{figure*}[t]
\centering
\includegraphics[width=16cm]{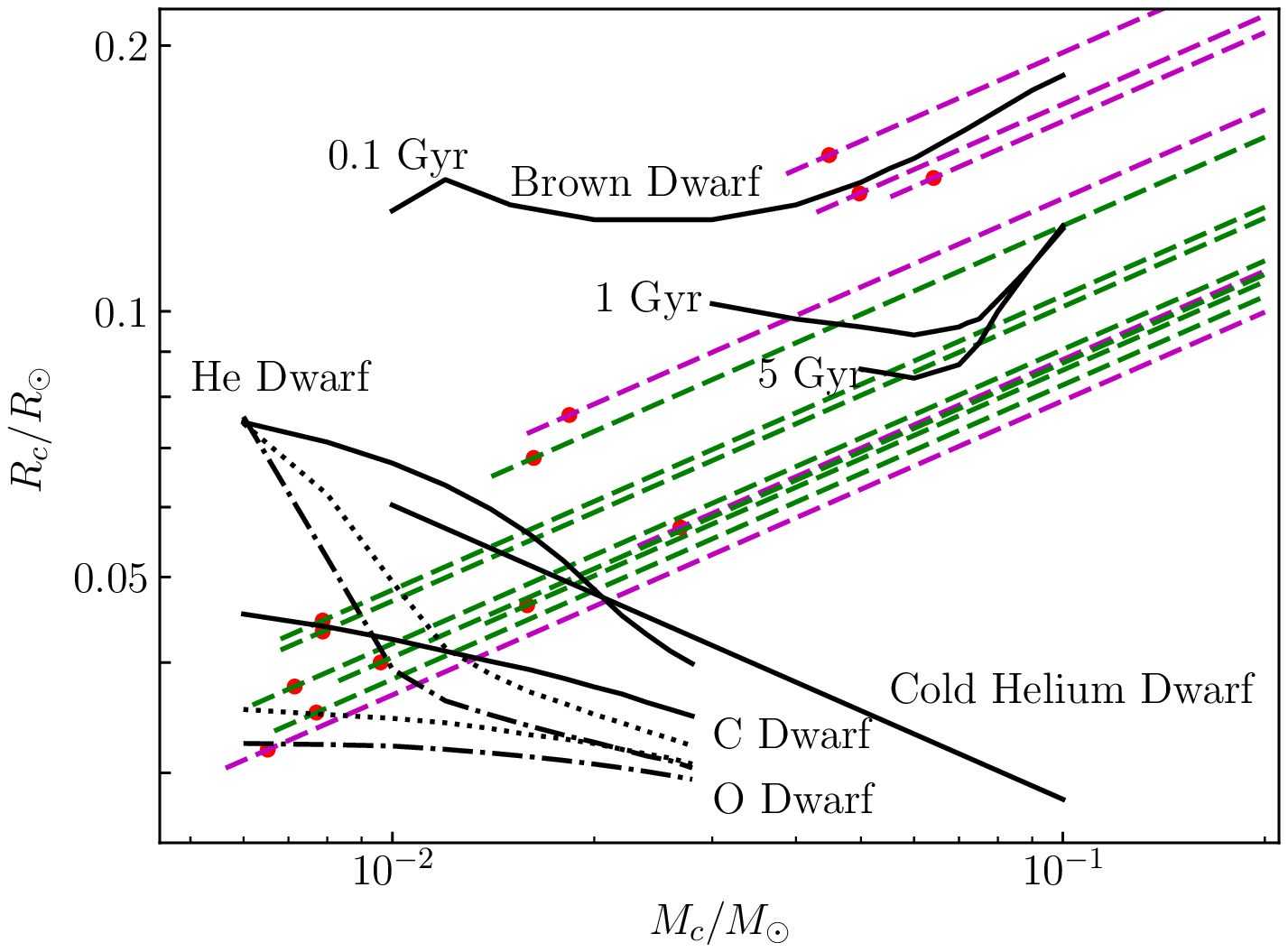}
\caption{Assumption of a Roche lobe-filling
companion implies a mass-radius relation 
$R_{\rm c} =
0.082\,(M_{\rm c}/0.1M_{\odot})^{1/3} (P_{\rm orb}/40\,\min)^{2/3}\, R_{\odot}$, 
%$R_{\rm lobe}=R_{\rm c} =1.515\times10^{-2}(M_{\rm c}/1M_{\odot})^{1/3}(P_{\rm orb}/1 {\rm   min})^{2/3}\, R_{\odot}$, 
  shown in logarithmic scale for different AMXPs with orbital periods around 40 min. From bottom up, the sources are IGR J17062$-$6143 \citep{Strohmayer18}, XTE J1807$-$294 \citep{Campana03}, XTE J1751$-$305 \citep{Papitto10}, XTE J0929$-$314 \citep{Galloway02}, MAXI J0911$-$655 \citep{Sanna17}, IGR J16597$-$3704 \citep{Sanna18b}, \psrtar\ \citep{krimm2007}, NGC\,6440\,X-2 \citep{Altamirano10}, IGR J17494$-$3030 \citep{Ng21}, HETE J1900.1$-$2455 \citep{Kaaret06}, IGR J17379$-$3747 \citep{Sanna18c}, SAX J1808.4$-$3658 \citep{Wijnands98}, and IGR J00291$+$5934 \citep{Galloway05}. 
   The magenta dashed lines show the AMXPs with detected type-I X-ray bursts, while the green dashed lines show the AMXPs without type-I X-ray bursts. The red points mark the inclination angle at 60$^{\circ}$. We note that from all AMXPs with the orbital period shorter than 1 hr, except MAXI J0911$-$655 and IGR J17494$-$3030, no type-I X-ray bursts were detected.
  We also show the mass-radius relations for low mass regime H-poor dwarf equation of states for O (dot-dash lines), C (dotted lines), and  He (solid lines) as well as low ($10^4$ K, lower lines) or high ($3\times 10^4$ K, upper lines) central temperatures.
  %the companion masses, $M_{\rm c}$, and radii, $R_{\rm c}$, plane, with low mass regime degenerate dwarf equation of states for different compositions O/C, and cold He $10^{4}$ K or higher $3\times10^{6}$ K central temperatures. 
 }
\label{fig:mc_rc}
\end{figure*}
%%%%%%%%%%%%%%%%%%%%%%%% Fig-9 %%%%%%%%%%%%%%%%%%%%%%%%

% suggested additions by dg
The non-detection of bursts in \psrtar\ is likely a consequence of a low burst rate coupled with a low observational duty cycle in the X-ray observations, such that any bursts that might occur would fall within data gaps.
The thermonuclear burst rate depends on the composition of the accreted fuel and the local mass accretion rate, $\dot{m}$, 
which, in turn, determine how much fuel must be accreted prior to ignition and the ignition depth. 
The known AMXPs can be divided into two distinct groups based on their orbital periods; either around 40 min or within the range of 2--11 h \citep[see e.g.,][]{Campana18}. Stellar evolution models predict that AMXPs with an orbital period longer than one hour should host a highly evolved, hydrogen-rich brown dwarf companion star, and the heating from steady burning of the accreted H fuel prior to ignition will lead to burst recurrence times of hours to days \citep[see e.g.,][]{galloway06,Heger07,falanga11,ferrigno2011,deFalcoa,deFalcob,ZLI2018,Kuiper20}. However, ``ultracompact'' X-ray binaries (UCXBs) with orbital periods 
$<80$~min are expected to have  low-temperature C, O or He white dwarf companions, that is, hydrogen-poor, highly evolved dwarfs \citep{deloye03}. The mass-radius relation of AMXPs are shown in Fig. \ref{fig:mc_rc},  by the assumption of a Roche lobe-filling companion \citep{Paczynski71}. The corresponding $M_{\rm c}$ versus $R_{\rm c}$ relations for different stellar evolution models are also provided in Fig.~\ref{fig:mc_rc}.  
% Accreting almost pure He or C, O matter composition, longer accretion time are predicted to have a substantial amount of H to trigger a short type-I X-ray burst, or larger ignition depth to have pure He bursts \citep[see e.g.,][]{galloway06}.
The expected burst recurrence time for such systems, even if reaching the same accretion rates during outbursts as the H-rich systems, are substantially lower, significantly reducing the chance of detecting bursts \cite[e.g.,][]{cumming03}.

We can roughly estimate the likely burst recurrence time based on the persistent flux of the source during the outburst
%
% what flux do you adopt for this calculation? 
% For F_bol = 2e-9 (at the peak), and assuming isotropy, I get about 7e3 g/cm^2/s
The local mass accretion rate per unit area onto the NS is $\dot{m}=4\pi d^2F_{\rm bol}(1+z)(4\pi R_{\rm NS}^2(GM_{\rm NS}/R_{\rm NS}))^{-1}$, i.e., $\dot{m}\approx3.7\times10^3 d^2_8~{\rm g~cm^{-2}~s^{-1}}$ and $\dot{m}\approx1.9\times10^3 d^2_8~{\rm g~cm^{-2}~s^{-1}}$ for the 2018 and 2019 outbursts, respectively. 
%
% I think the ignition column adopted below is a little high, cumming03 suggests y~1e9. I don't see 
% the value of 1e10 in the cited in 't Zand paper...?
If the accreted matter is pure helium, the expected recurrence time for type-I X-ray bursts is $t_{\rm rec}\approx y/\dot{m}\approx116~{\rm d}~(y/10^{10}~{\rm g~cm^{-2}})(\dot{m}/10^3~ {\rm g~cm^{-2} ~s^{-1}})^{-1}$, where $y$ is the ignition depth of a helium burst \citep[see e.g.,][]{Intzand07}. 
For \psrtar,\ during its 2018 and 2019 outbursts, we estimate a burst recurrence time at the peak of the outburst of 30~d, 
%
% this value is based on the 1e10 g/cm^2 column, and the 3.7e3 g/cm^2/s accretion rate
% if you adopt instead 7e3 (as calculated above), the recurrence time goes down to ~15 d, but the
% conclusions are unchanged --- dkg
%
longer than the total duration of the outbursts. If the accreted matter is composed of C and O, the  X-ray bursts would be expected to occur at even larger ignition depths. We expect that the recurrence time should be much longer. 
Given recurrence times in excess of the outburst duration, it is questionable whether sufficient material would accrete to produce even a single burst.

While no type-I burst events were detected during outbursts of transient ultracompact AMXPs, including \psrtar, 
we note that some UCXBs 
% with a hydrogen-deficient donor star (degenerate O, C, or He dwarfs star) 
can sustain  accretion persistently and exhibit type I bursts \citep[e.g][]{cumming03,Intzand07,falanga08}. This property may explain the recent detection of a presumed type I X-ray burst in the AMXP MAXI~J0911$-$655 \citep{ATel13754,ATel13760}, and two intermediate duration X-ray bursts in IGR J17062$-$6143 \citep{Degenaar13,Keek17}. The transient AMXPs spend most of the time in a quiescent phase, with X-ray luminosities of $10^{31}-10^{33}$~erg~s$^{-1}$ corresponding to accretion rates that are likely far too low to produce thermonuclear bursts.
% and sometimes undergo X-ray outbursts reaching luminosities of $10^{36}-10^{37}$ erg s$^{-1}$ have no physical conditions to 

%(see Fig. \ref{fig:mc_rc})   The assumption of a Roche lobe-filling companion \citep{Paczynski71} implies the mass-radius relation shown in Fig. \ref{fig:mc_rc}) (dashed lines), and the corresponding $M_{\rm c}$ versus $R_{\rm c}$ relations for different stellar evolution models are also shown in Fig.~\ref{fig:mc_rc}. For a mean inclination of $i \sim$  60\degr\ (red dot points) only low temperature C/O dwarfs are allowed, may excluding He dwarfs which would require very low \red{inclinations $i \leq$  ??? \degr.} 
 %Given the mass function, $f(M)$= ($M_{\rm c}\, \sin\,i)^{3}$/($M_{\rm c}$ + $M_{\rm NS})^{2}$, an important free parameter is the inclination angle of the system.

%On the other hand, there were no X-ray eclipses or dips detected in the light curves, putting an upper limit on the binary inclination of $i < 80\degr-85\degr$ for a Roche lobe-filling companion. Given the mass function, $f(M)$= ($M_{\rm c}\, \sin\,i)^{3}$/($M_{\rm c}$ + $M_{\rm NS})^{2}$, an important free parameter is the inclination angle of the system. This strongly constrains the evolution of the system.

\section{Summary}
\label{sect_disc}

In this work, we analyze all public high-energy data and report the outburst profiles and the spectral and timing properties of  \psrtar\ observed by \Integ, \xmm, \nustar, \nicer, and \hxmt, during its  2018 and 2019 outbursts. We found these two outbursts showed quite similar behavior in several aspects. The outburst profiles showed a similar shape, which can be explained by the disk instability models. The broadband spectra -- in the energy range of 1--250 keV for the brighter 2018 outburst and in the range of 1.5--80 keV for the fainter 2019 outbursts -- were well fitted by the thermal Comptonization model \compps\ with a similar set of parameters: the electron temperature $kT_{\rm e}=40$--50 keV, Thomson optical depth $\tau\sim 1.3$, and blackbody seed photon temperature $kT_{\rm bb,seed}\sim $0.7--0.8~keV. 
The spectral shape is very similar to those observed in many other AMXPs \citep{papitto20}.

We performed a coherent timing analysis for the 2018 and 2019 outbursts together. Pulsed emission has been detected from \psrtar\ in the energy range of 1--10 keV  using \nicer\ and \xmm\ data, in 3--60 keV using \nustar\ during both outbursts, and in the 5--100 keV range using \hxmt\ during the 2018 outburst. 
We detected an increase of the pulse fraction from 4\%  to  7.5\% from 1 to 5 keV saturating at higher energies.
We found no evidence for a change in the spin frequency from our data set. Comparing the observed times of ascending node passage with the predicted values, we concluded that the binary system had a constant orbital period since the first outburst in 2007.   

\psrtar\ is the third LMXB that has exhibited pairs of closely-spaced outbursts, which (assuming a steady accretion rate onto the disk) must arise from incomplete accretion of the fuel accumulated in the disk, during the first outburst \cite[]{hartman11}. 
Those authors explained the outburst pairs observed in IGR~J00291+5934 as the result of a ``quasi-propeller'' regime, which shut off the accretion before the disk was exhausted. There are a number of differences between the outburst pairs in the two systems; in \psrtar\ the pairs of outbursts were separated by 2.1 and 1.2~yr, whereas for the other systems, it was 30~d. IGR~J00291+5934 has a brown dwarf donor, whereas \psrtar\ probably has a He WD donor. Hence, the size and composition of the disk are very different in these two systems. Furthermore, IGR~J00291+5934 rotates at 599~Hz, substantially faster than \psrtar\ at 182~Hz; the other system exhibiting outburst pairs, XTE~J1118+480, has a black hole primary.

While an assessment of the outburst behavior in \psrtar\ in the context of the disk instability model is beyond the scope of this paper, the accumulated outburst history of the three sources that have been observed to exhibit such double outbursts appears to offer an excellent opportunity for a more in-depth and comprehensive comparison, which may provide significant new insights into the disk instability mechanism.

\begin{acknowledgements}
We thank the anonymous referee for valuable comments.
ZL thanks the International Space Science Institute in Bern for the hospitality. This work is supported by the National Key R\&D Program of China (2016YFA0400800).
ZL was supported by National Natural Science Foundation of China  (U1938107, 11703021, U1838111, 11873041), and Scientific Research Fund of Hunan Provincial Education Department (18B059). 
JP and SST were supported by the grant 14.W03.31.0021 of the Ministry of Science and Higher Education of the Russian Federation and the Academy of Finland grants 317552, 322779, 324550, 331951, and 333112. 
SZ and SNZ were supported by National Natural Science Foundation of China (Nos. U1838201 and U1838202).
This research has made use of data obtained from the High Energy Astrophysics Science Archive Research Center (HEASARC), provided by NASA's Goddard Space Flight Center, and also from the HXMT mission, a project funded by the China National Space Administration (CNSA) and the Chinese Academy of Sciences (CAS). 
\end{acknowledgements}

\bibliographystyle{aa}
\bibliography{pulsars}

\end{document}